\documentclass[aps,prl,twocolumn,showpacs,floatfix,superscriptaddress,longbibliography]{revtex4-1}
\pdfoutput=1 

\usepackage{amssymb,amsmath,amstext}                
\usepackage{graphicx}                                               
\usepackage{epstopdf}                                               
\usepackage{color}                                                     
\usepackage{bm}                                                        
\usepackage{appendix}                                              
\usepackage[utf8]{inputenc}
\usepackage{bbold}
\usepackage{bbm}   
\usepackage{ulem}
\usepackage{latexsym}
\usepackage[colorlinks=true,allcolors=blue]{hyperref}
\usepackage{mathtools}

\def\be{\begin{equation}}
\def\ee{\end{equation}}
\def\bea{\begin{eqnarray}}
\def\eea{\end{eqnarray}}

\def\bi{\begin{itemize}}
\def\ei{\end{itemize}}
\def\ben{\begin{enumerate}}
\def\een{\end{enumerate}}

\usepackage[dvipsnames]{xcolor}
\definecolor{dgreen} {RGB}{0,100,0}

\newcommand{\iu}{\mathrm{i}\mkern1mu}

\newcommand{\diff}{{\mathrm{d}}}

\newcommand{\definition}{\coloneqq}

\let\vec\mathbf

\allowdisplaybreaks

\begin{document} 

\title{Dissipationless Vector Drag---Superfluid Spin Hall Effect}
\author{Andrzej Syrwid} 
 \affiliation{Department of Physics, KTH Royal Institute of Technology, SE-106 91 Stockholm, Sweden}

 \author{Emil Blomquist}
 \affiliation{Department of Physics, KTH Royal Institute of Technology, SE-106 91 Stockholm, Sweden}

 \author{Egor Babaev}
 \affiliation{Department of Physics, KTH Royal Institute of Technology, SE-106 91 Stockholm, Sweden}

\begin{abstract}
Dissipationless flows in single-component superfluids have a  significant degree of universality. In \textsuperscript{4}He, the dissipationless mass flow occurs with a superfluid velocity determined by the gradient of the superfluid phase.
However, in interacting superfluid mixtures, principally new effects appear.
In this Letter, we demonstrate a new kind of dissipationless phenomenon arising in mixtures of interacting bosons in optical lattices. We point out that for a particular class of optical lattices, bosons condense in a state where one of the components' superflow results in dissipationless mass flow of the other component, in a direction different from either of the components' superfluid velocities. The free-energy density of these systems contains a vector product-like interaction of superfluid velocities, producing the dissipationless noncollinear entrainment.
The effect represents a superfluid counterpart of the Spin Hall effect.
\end{abstract}

\date{\today}


\maketitle

In 1975 Andreev and Bashkin demonstrated the principally new dissipationless drag-transport effect in an interacting superfluid mixture~\cite{AndreevBashkin1975}.
Namely, they showed that a nonzero superfluid velocity of one component induces a collinear dissipationless mass transfer of the other component.
This entrainment effect is present in superfluids, superconductors---including those with unconventional pairing~\cite{Leggett1975}---and dense nuclear matter~\cite{SJOBERG1976,Chamel2008}.
It  determines observed dynamics of pulsars~\cite{Alpar1984,Alford2008,Babaev2009neutron}
and can cause phase transitions to new types of superfluids~\cite{Kuklov2003,Kuklov20042,Kuklov2004,Kuklov2006,Herland2010,Dahl2008}.

As shown in~\cite{AndreevBashkin1975}, in the presence of intercomponent interactions, the free-energy density describing a binary superfluid mixture should necessarily include  a scalar product of the superfluid velocities $ \mathbf v_a $ and $ \mathbf v_b $, i.e.,
$ f = \rho_a v_a^2 / 2 + \rho_b v_b^2 / 2 + \rho_{ab}\,{\bf v}_{a} \cdot {\bf v}_{b}$.
The resulting superflow $ {\bf j}_a = \partial f / \partial {\bf v}_a = \rho_a {\bf v}_a+\rho_{ab} {\bf v}_b $ indicates that even if $ {\bf v}_a = \mathbf 0 $, there will still be a nonzero superflow of component $ a $ with the superfluid velocity $ \vec v_b $. 

In the condensed matter context, the effect became of great interest with the advent of optical lattices, which allow for precise control of strongly correlated superfluids~\cite{greiner2002quantum,bloch2005ultracold}. The strength of the Andreev-Bashkin drag is controlled by the optical lattice parameters in combination with on-site interactions~\cite{Sellin2018}. It was shown that the effect, in relative terms, can be arbitrarily strong and that the  Andreev-Bashkin drag coefficient $ \rho_{ab} $ can also become negative. In the latter case, one deals with a counterflow, where the flow of one component generates a mass flow of the other component in the opposite direction~\cite{Kuklov2003,Kuklov20042,Kuklov2004,Kuklov2006}.
It was pointed out that the effect should lead to the formation of new superfluid states where only dissipationless coflow (paired superfluids) or only counterflow (supercounterfluids)  can exist~\cite{Kuklov2003,Kuklov20042,Kuklov2004,Kuklov2006,soyler2009sign,Dahl2008,Herland2010,Sellin2018,EgorBook}. 
At the same time, even relatively weak drag substantially changes rotational responses~\cite{Dahl2008_2,Dahlprl2018}.

In this Letter, we demonstrate the existence of a new dissipationless phenomenon in superfluid mixtures, where the dissipationless superfluid entrainment is not collinear with superfluid velocities. 
Namely, we show that the superflow-superflow interaction
can, in general, be described by a nontrivial tensor $  \rho_{\alpha\beta}^{ij}$ which enters the bilinear free-energy density:
\begin{equation}
    \label{Fvelocities}
        f
        =
        \frac{1}{2}
        \sum_{\alpha\beta}
        \sum_{ij}
        \rho_{\alpha\beta}^{ij}
        v_{\alpha}^i v_{\beta}^j
        \,.
\end{equation}
Here Greek subscripts and Roman superscripts, respectively, label components and Cartesian directions of the superfluid velocity vector $ v_\alpha^i $ and the superfluid stiffness tensor $ \rho_{\alpha\beta}^{ij} =\rho_{\beta \alpha}^{ji} $. 
The latter describes both kinetic ($ \alpha = \beta $) and drag ($ \alpha \neq \beta $) phenomena.
While physical properties of the system are encoded in the tensor $  \rho_{\alpha\beta}^{ij}$,  as we elaborate below, a direct interpretation of individual coefficients may be deceptive since they depend on the choice of the coordinate system.

In this Letter we consider a two-dimensional two-component system---$ i,j \in \{x, y\} $ and $ \alpha, \beta \in \{a, b\} $---and study the drag-related elements $ \rho_{ab}^{ij} $.
In such a case, the quantities 
\begin{align}
    \rho_\parallel
    =
    \left(\rho_{ab}^{xx}+\rho_{ab}^{yy}\right) / 2
    \,, 
    \qquad
    \rho_\perp
    =
    \left(\rho_{ab}^{xy}-\rho_{ab}^{yx}\right) / 2
    \,,
    \label{DeltaSigma}
\end{align}
are coordinate system independent~\cite{SI}.
All other pairwise combinations vary under rotation, and it turns out that it is always possible to find a Cartesian coordinate system in which $\rho_{ab}^{xy}+\rho_{ab}^{yx} = 0$.
If, in addition, the difference $ \rho_{ab}^{xx}-\rho_{ab}^{yy}$ is negligible---which in principle can be guaranteed in certain situations---one finds
\begin{equation}
    \label{fDelta}
    f
    =
     f_{0}
    +
    \rho_\parallel \, {\bf v}_a \cdot {\bf v}_b
    +
    \rho_\perp \, (v_a^x v_b^y - v_a^y v_b^x)
    \,,
\end{equation}
where $ f_{0} = \sum_\alpha\sum_{ij}\rho_{\alpha\alpha}^{ij}v_{\alpha}^i v_{\alpha}^j/2  $  represents the standard kinetic contribution to the free-energy density. The corresponding superflows read
\begin{equation}
\begin{split}
    &{\bf j}_{a}
    =
    {\bf j}_{0a}
    +\rho_\parallel \, {\bf v}_{b}
    +\rho_\perp \, ( v_{b}^{y}{\bf e}_x-v_b^x{\bf e}_y) \,,
    \\
    &{\bf j}_b
    =
    {\bf j}_{0b}
    +\rho_\parallel \, {\bf v}_{a}
    -\rho_\perp \, ( v_{a}^{y}{\bf e}_x-v_a^x{\bf e}_y) \,,
\end{split}
\end{equation}
where $ {\bf j}_{0\alpha} = \partial f_{0} / \partial {\bf v}_{\alpha} $ and $ {\bf e}_i $ denotes the unit vector in the $ i$\textsuperscript{th} direction.
The existence of the vector product-like contribution to $ f $ given by $ \rho_\perp $  constitutes a novel superfluid effect where a nonzero superflow of one component induces a perpendicular superflow response of the other component. 
This new phenomenon, which we coin  {\it vector drag}, may be viewed as an intercomponent-interaction-driven counterpart of the Spin Hall effect~\cite{d1971possibility} originating in the spin-orbit coupling, discussed  particularly in hybrid structures involving  superconductors \cite{Bergeret2016,Linder2017}.

Let us now provide both analytical and numerical evidence for the existence of a nonzero $ \rho_\perp  $, and thus the existence of the vector-drag phenomenon. Our starting point in deriving the effective model in Eq.~\eqref{Fvelocities} is a two-component Bose-Hubbard-type model on a rectangular lattice, given by the Hamiltonian
\be
\label{eq:Hamiltonian}
    \hat{H}
    =
    -
    \sum_{\alpha}
    \sum_{ij}
    t_{ij\alpha} \, \hat b_{i\alpha}^\dagger \hat b_{j\alpha}
    +
    \frac{1}{2}
    \sum_{\alpha\beta}
    \sum_{ij}
    U_{ij\alpha\beta} \, \hat n_{i\alpha} \hat n_{j\beta} \,.
\ee
Here $ \hat b_{i\alpha} $ ($ \hat b^\dagger_{i\alpha} $) denotes the bosonic annihilation (creation) operator of component $ \alpha $ at site $ i $, and $ \hat n_{i\alpha} = \hat b^\dagger_{i\alpha} \hat b_{i\alpha} $ is the corresponding particle number operator.
The mass of the $ \alpha $-component boson is $ m_\alpha $, and the
rectangular lattice
consists of $ N \times N $ sites---with lattice vectors $ \mathbf a_x = l_x \mathbf e_x $, $ \mathbf a_y = l_y \mathbf e_y $ and lattice constants $ l_x $, $ l_y $---on which we impose periodic boundary conditions.
For simplicity, we assume constant particle number densities $n_\alpha$ and restrict ourselves to on-site and nearest-neighbor interactions whilst allowing for nearest-neighbor and next-nearest-neighbor hopping. 
The problem is amenable to analytical treatment only in the weakly interacting regime.
Here, like in the case of the Andreev-Bashkin effect~\cite{Fil&Shevchenko2005,Linder&Sudbo2009,Hofer1012,Hartman2018,nespolo2017andreev}, the drag effects are expected to be inherently small.
We will first demonstrate the existence of vector drag in the weakly interacting regime analytically.
Then we investigate the effect in the strongly correlated regime by employing large-scale quantum Monte-Carlo calculations.

\begin{figure}[t]
\includegraphics[width=1\columnwidth]{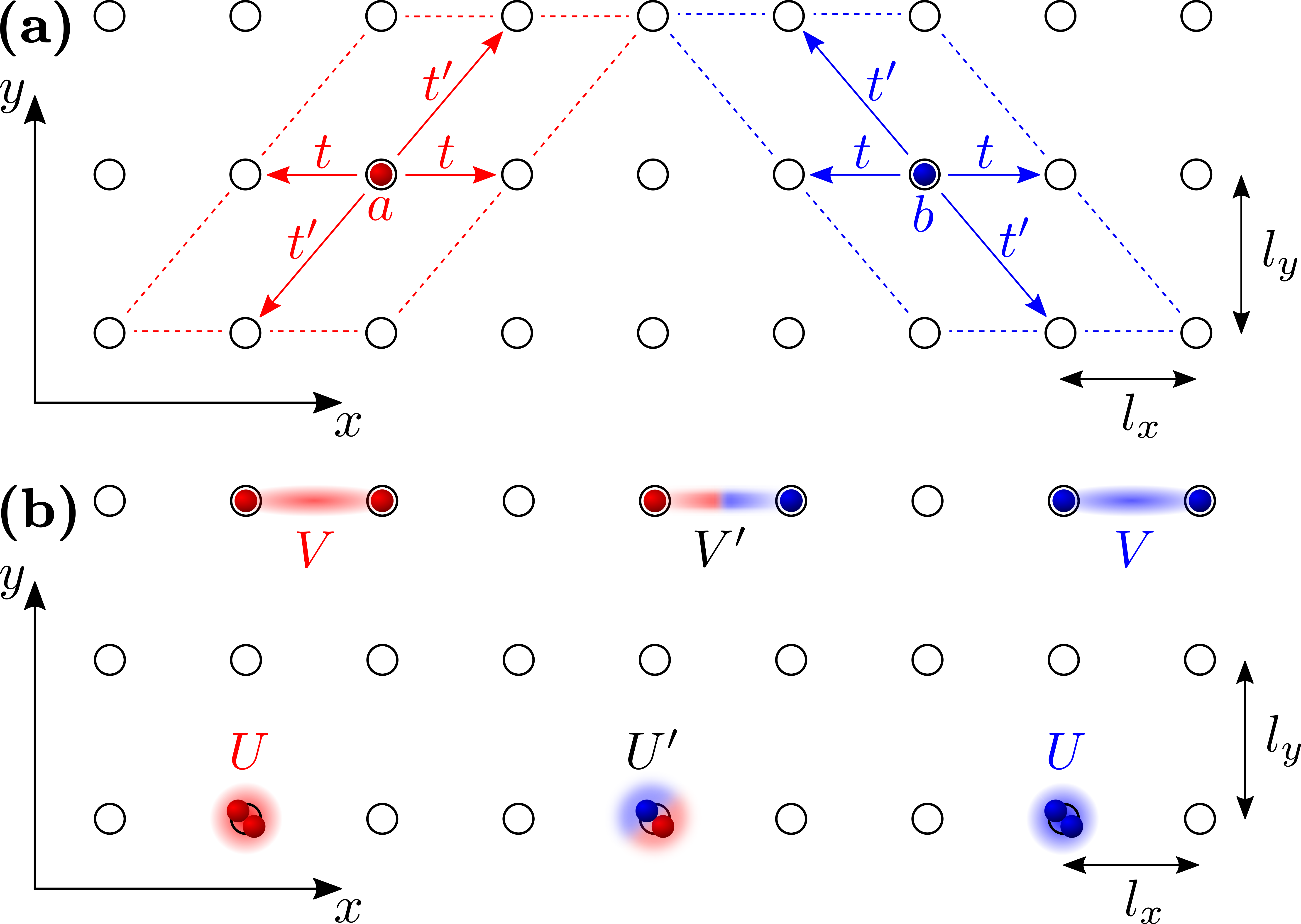}
\caption{
Illustration of model parameters.
(a) The allowed hopping directions.  Both components can hop in the $x$ direction with the amplitude $t$. In addition, component $ a $ (red) and component $ b $ (blue) can hop along the diagonal and antidiagonal, respectively. The latter process is characterized by the amplitude $t'$. Intra- and intercomponent interactions are schematically illustrated in (b). 
}
\label{fig:model}
\end{figure}

The standard Andreev-Bashkin effect can be analytically calculated in macroscopic weakly interacting systems.
That was previously done for square and triangular lattices and in a continuum~\cite{Fil&Shevchenko2005,Linder&Sudbo2009, Hofer1012, Hartman2018}.
We begin by employing a similar analytic approach to establish the new phenomenon: the vector drag.
Consider a weakly interacting regime where at low enough temperatures both components are condensed into the zero-momentum mode. In such a case, the Hamiltonian \eqref{eq:Hamiltonian} can be approximated and subsequently diagonalized in momentum space. 
The corresponding zero-temperature free-energy density \eqref{Fvelocities}   is then obtained as the ground state energy.

As discussed above, the vector drag should be the most transparent when $ \rho_{ab}^{xy} = -\rho_{ab}^{yx} $, which is guaranteed for systems being invariant under a reflection in either of the two lattice vectors combined with an exchange of components $ a \leftrightarrow b $~\cite{SI}.
Our aim is to construct a microscopic model which exhibits $ \rho_\perp \neq 0 $ in addition to satisfying the above-mentioned symmetry.
A simple choice of parameters obeying these conditions is $m_a=m_b=m$, $n_a=n_b=n$, and
\begin{align}
    \label{eq:parameter_choice}
    \begin{split}
        t_{ij\alpha}
        &=
        \left\{
        \begin{array}{ll}
            t  & \,\,\, \text{for} \;\; 
             {\bf r}_i = {\bf r}_j  \pm {\bf a}_x  \\
            t' & \,\,\, \text{for} \;\; 
             {\bf r}_i = {\bf r}_j  \pm {\bf a}_x 
            \pm (\delta_{a\alpha} - \delta_{b\alpha}) {\bf a}_y \\
            0  & \,\,\, \text{otherwise}
        \end{array}
        \right. \!\!,
        \\
        U_{ij\alpha\beta}
        &=
        \left\{
        \begin{array}{ll}
            U \delta_{\alpha\beta} + U' (1-\delta_{\alpha \beta }) & \,\,\, \text{for} \;\;   {\bf r}_i = {\bf r}_j 
            \\
            V \delta_{\alpha \beta} + V' (1-\delta_{\alpha \beta }) & \,\,\, \text{for} \;\; 
             {\bf r}_i = {\bf r}_j  \pm {\bf a}_x  
            \\
            0  & \,\,\, \text{otherwise}
        \end{array}
        \right. \!\!,
        \end{split}
\end{align}
where ${\bf r}_i$ indicates the $i$\textsuperscript{th} lattice site position. The resulting model is illustrated in Fig.~\ref{fig:model}.
While in general $\rho_{ab}^{xx}\neq \rho_{ab}^{yy}$, it turns out that when both $ \rho_{ab}^{xx} $ and $ \rho_{ab}^{yy} $ are nonzero and of the same sign---which for the considered parameter region is the case---one can in principle completely eliminate $ \rho_{ab}^{xx}-\rho_{ab}^{yy} $. This is achieved by rescaling the ratio $ l_x/l_y $ by $ \sqrt{\rho_{ab}^{yy} / \rho_{ab}^{xx}} $ while at the same time keeping all other model parameters fixed, which leaves the $ l_x,l_y $-independent $ \rho_\perp $ unchanged~\cite{SI}.
In this way we can realize the model effectively described by Eq.~(\ref{fDelta}).
We find that for the lattice geometry illustrated in Fig.~\ref{fig:model}(a), this rescaling leads to $ l_x < l_y $ in the region of interest, which is consistent with keeping the nearest-neighbor interactions in the $x$ direction only---typical interatomic interactions rapidly decay with increased separation distance.
Nevertheless, omitting the nearest-neighbor interactions along the $y$ direction is merely a simplification to reduce the number of system parameters and does not change the main result. Namely, we would like to stress that the presence of vector drag is not limited to systems having nearest-neighbor interaction in one direction only.
In what follows we will restrict ourselves to $m=1$, $n=1/2$, $ U = 1 $, $ U' = 0.9 $, and $ V' = 0.9 V $.

\begin{figure}[t]
    \includegraphics[width=1\linewidth]{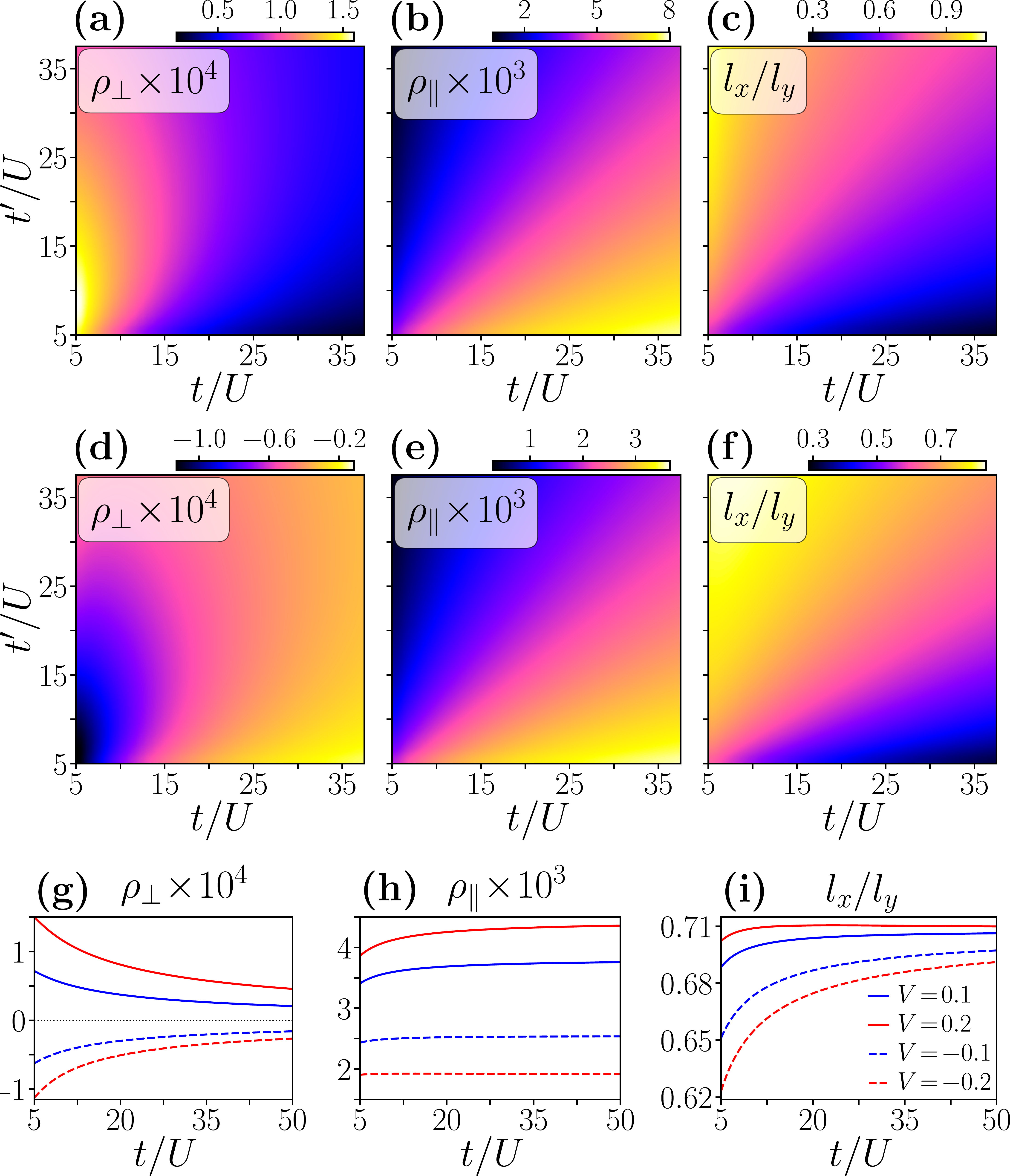}
    \caption{
        Vector drag in the weakly interacting regime.
        The left column (a), (d), (g) presents the $ l_x,l_y $-independent vector-drag coefficient $ \rho_\perp $. The middle column (b), (e), (h) shows the $ l_x/l_y $-dependent collinear-drag coefficient $ \rho_\parallel $, computed using the ratio $ l_x/l_y $  plotted in the right column (c), (f), (i), which is adjusted such that $ \rho_{ab}^{xx}=\rho_{ab}^{yy} $ required for the system to be described by Eq.~(\ref{fDelta}).
        The upper (a)-(c) and middle (d)-(f) row display how the vector drag $\rho_{\perp}$,  Andreev-Bashkin drag $\rho_{\parallel}$, and $l_x/l_y$ depend on $ t/U $ and $ t'/U $ when $ V = 0.2 $ and $ V = -0.2 $, respectively. 
        Even though the presence of the nearest-neighbor interactions modifies the amplitude of $\rho_{\parallel}$, it does not change its character determined by the sign.
         On the contrary, both the sign and magnitude of the vector drag strongly depend on the sign and magnitude of $V$. Note that the sign change of $V$ affects the ratio $\rho_{ab}^{yy}/\rho_{ab}^{xx}$ implying
        differences between $l_x/l_y$ visible in (c) and (f). 
        The same quantities, $\rho_\perp,\rho_{\parallel}$ and $l_x/l_y$, but for $ t/U = t'/U $ and $V=\pm0.1,\pm0.2$, are presented in the bottom row (g)-(i).
    }
    \label{dragsweak}
\end{figure}

Our findings indicate that in the weakly interacting regime, the macroscopic system harbors a substantial vector drag when $ V \neq 0 $, at least within our approach.
In Fig.~\ref{dragsweak} we present 
the analytically derived drag coefficients $\rho_\perp$ and $\rho_\parallel$  versus $ t/U $ and $ t'/U $, for different values of $V$, in panels  (a), (d), and (g)
and in  (b), (e), and (h), respectively.
The latter quantity is calculated for $l_x/l_y$ adjusted such that $\rho_{ab}^{xx}=\rho_{ab}^{yy}$, which entails $l_x<l_y$ apart from the region where $ t' \gg t $, see panels (c), (f), and (i) of Fig.~\ref{dragsweak} .
Since both the magnitude and sign of $\rho_\perp$ are determined by the magnitude and sign of $V$, the nearest-neighbor interactions are important for the vector-drag phenomenon in the considered regime.

\begin{figure}[t]
  \centering
  \includegraphics[width=1\columnwidth]{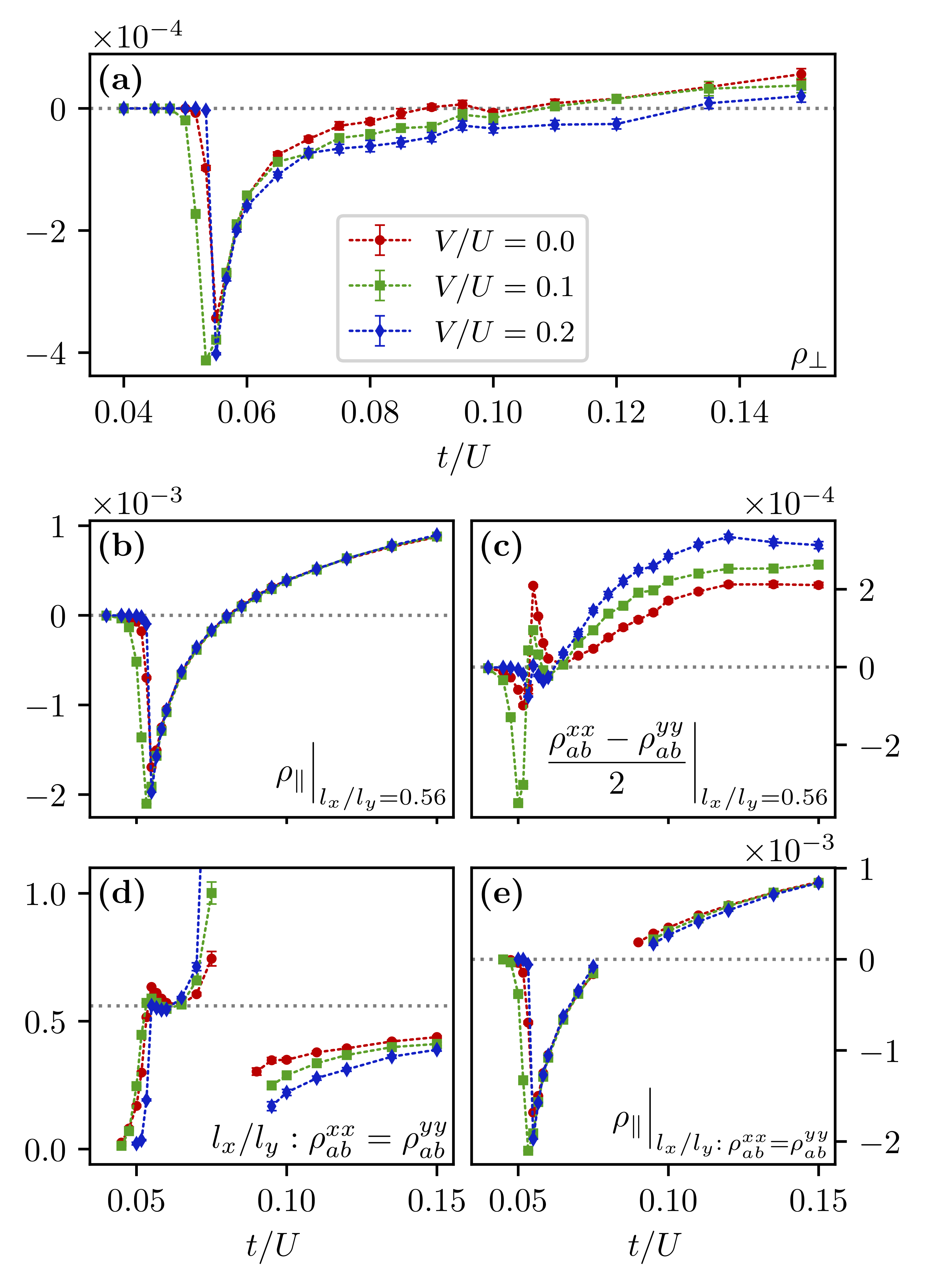}
  \caption{
    Vector drag in the strongly interacting regime.
    The $l_x,l_y$-independent vector-drag coefficient $\rho_\perp $ is presented in panel (a), while the $l_x/l_y$-dependent quantities $\rho_\parallel$ and $(\rho_{ab}^{xx}-\rho_{ab}^{yy})/2$ are shown in panels (b) and (c), here using $l_x/l_y=0.56$.
    An alternative ratio $ l_x/l_y $---for which $ (\rho_{ab}^{xx} - \rho_{ab}^{yy})/2 $ is completely eliminated---and the corresponding $\rho_{\parallel}$ are plotted in panels (d) and (e), respectively.
    In the latter panels, the missing data points around $ t/U \approx 0.08 $ are due to $ \rho_{ab}^{xx} $ and $ \rho_{ab}^{yy} $ having opposite signs.
    All results were obtained using $ V/U = 0, 0.1, 0.2 $, $ t/U = t'/U $, $ N = 10 $, and  $ \beta = N / t $.
    }
  \label{fig:drag_coeffs_strong}
\end{figure}

After demonstrating the effect analytically in the weak-coupling regime, we proceeded to study the model, Eqs.~(\ref{eq:Hamiltonian}) and (\ref{eq:parameter_choice}), deep inside the strongly correlated regime.
For this purpose, we performed large-scale worm-algorithm Monte Carlo simulations~\cite{Prokofev1998,PhysRevB.75.134302}.
Using a generalization of the Pollock-Ceperley formula~\cite{SI}, which in its original form allows computing the superfluid density through winding number statistics~\cite{PhysRevB.36.8343}, we obtained the drag-related elements of the more general tensor:
\begin{equation}
    \label{eq:Pollock-Cepereley-rectangular}
    \rho_{ab}^{ij}
    =
    \frac{1}{\beta}
    \frac{l_i l_j}{l_x l_y}
    \langle w_a^i w_b^j \rangle \,.
\end{equation}
Here $ \beta $ is the inverse temperature and $ \langle w_a^i w_b^j \rangle $ are winding number correlations---the winding number $ w_\alpha^i $ is the net number of times $ \alpha $-type particles cross the periodic boundary along the $i$\textsuperscript{th} direction.
Similarly as in the weakly interacting case, we consider the system with constant particle densities equal for both components, i.e., $n_a=n_b=n=1/2$.

The results obtained for the parameters $ N = 10 $, $ t = t' $, $ \beta = N / t $, and $ V/U = 0, 0.1, 0.2 $ are presented in Fig.~\ref{fig:drag_coeffs_strong}.
Here $ \rho_\perp $, $ \rho_\parallel $, and $ (\rho_{ab}^{xx} - \rho_{ab}^{yy})/2 $ are plotted against $ t/U $ in panels (a)-(c), respectively. At small values of $ t/U $, the system is in an insulating phase which is evident from the coefficients being identically zero. Then, at $ t/U \approx 0.053 $, the system undergoes a transition into a superfluid state with negative drag coefficients. By further increasing $ t/U $, $ \rho_\perp $ and $ \rho_\parallel $ increase in value and eventually change sign, whereas $ (\rho_{ab}^{xx} - \rho_{ab}^{yy})/2 $ reveals a more complex dependence on $ t/U $. The behavior of the  Andreev-Bashkin  drag $ \rho_\parallel $ is very similar to that of the conventional two-dimensional two-component Bose-Hubbard model \cite{Sellin2018}.
Note that the magnitudes of both $\rho_{\parallel}$ and $\rho_{\perp}$ are maximal in the strongly correlated regime very close to the phase transition point, 
which is similar to the fact that in the previously studied systems the Andreev-Bashkin drag is also maximal close to insulating phases \cite{Sellin2018}.  

For $ l_x/l_y = 0.56 $, the difference $ (\rho_{ab}^{xx} - \rho_{ab}^{yy})/2 $ is comparable to $ \rho_\perp $ in magnitude and should not be ignored.
Nevertheless, by adjusting the $ l_x/l_y $ ratio as previously mentioned we can prompt $ \rho_{ab}^{xx} = \rho_{ab}^{yy} $ in situations where $ 0 < \rho_{ab}^{yy} / \rho_{ab}^{xx} < \infty $.
This $ l_x/l_y $ ratio, as a function of $ t/U $, is shown in panel  (d) of Fig.~\ref{fig:drag_coeffs_strong}, and the resulting $ \rho_\parallel $ can be found in panel (e).
In the vicinity of $ t/U \approx 0.08 $ the ratio $ \rho_{ab}^{yy}/\rho_{ab}^{xx} $ does not fulfill the above-mentioned criteria which is why there are missing data points.

The analytical result for the weakly interacting macroscopic limit, and the unbiased numerical data for the strongly correlated regime, both demonstrate the new type of dissipationless transport manifested in the nonzero value of the vector-drag coefficient $ \rho_\perp $.
In the strong coupling regime, substantial nonzero
vector drag $ \rho_\perp $  exists  in case  of exclusively on-site interactions, i.e., for $ V = 0 $, for the system size considered.
The vector-drag coefficient's dependence on the system size with purely on-site interactions is further discussed in the Supplemental Material~\cite{SI}, where it is demonstrated that the effect is still present also in larger systems.

In summary, we demonstrated a new dissipationless transport phenomenon in superfluid mixtures in optical lattices.
Namely, we have shown that for a class of optical lattices, the free-energy density of an interacting superfluid mixture should contain a vector product-like interaction between the superfluid velocities.
This implies that in such a mixture the superflow of each component is not collinear with either of the components' superfluid velocity, i.e., a superflow of one of the components induces a  superflow of the other component in the orthogonal direction, in addition to the standard drag.
It may be viewed as an intercomponent-interaction-induced superfluid counterpart of the Spin Hall effect.

The strength of the effect was investigated both analytically in the weakly interacting macroscopic limit and numerically in the strongly correlated regime using large-scale worm-algorithm quantum Monte-Carlo simulations.
One way to realize this model is through a bilayer optical lattice loaded with dipolar bosons \cite{PhysRevA.90.043623, Safavi_Naini_2013}. 

The effect should also be present in multicomponent superconductors.
There, at the level of the Ginzburg-Landau model, it should 
manifest through the presence of  terms in the form of vector-like product of components of supercurrents,  i.e., mixed terms fourth order in fields and second order in gradients.
Investigation of these aspects will be presented in  future studies.

\section*{Acknowledgements}
E.Bl. and E.Ba. were supported by the Swedish Research Council Grants No. 2016-06122 and No. 2018-03659, and G\"{o}ran Gustafsson Foundation for Research in Natural Sciences.
A.S and E.Ba. acknowledge the support from Olle Engkvists stiftelse.
The computations were enabled by resources provided by the Swedish National Infrastructure for Computing (SNIC) at the National Supercomputer Centre (NSC) partially funded by the Swedish Research Council through Grant No. 2018-05973.
\bibliographystyle{apsrev4-1}
\bibliography{refs}

\begin{thebibliography}{34}%
\makeatletter
\providecommand \@ifxundefined [1]{%
 \@ifx{#1\undefined}
}%
\providecommand \@ifnum [1]{%
 \ifnum #1\expandafter \@firstoftwo
 \else \expandafter \@secondoftwo
 \fi
}%
\providecommand \@ifx [1]{%
 \ifx #1\expandafter \@firstoftwo
 \else \expandafter \@secondoftwo
 \fi
}%
\providecommand \natexlab [1]{#1}%
\providecommand \enquote  [1]{``#1''}%
\providecommand \bibnamefont  [1]{#1}%
\providecommand \bibfnamefont [1]{#1}%
\providecommand \citenamefont [1]{#1}%
\providecommand \href@noop [0]{\@secondoftwo}%
\providecommand \href [0]{\begingroup \@sanitize@url \@href}%
\providecommand \@href[1]{\@@startlink{#1}\@@href}%
\providecommand \@@href[1]{\endgroup#1\@@endlink}%
\providecommand \@sanitize@url [0]{\catcode `\\12\catcode `\$12\catcode
  `\&12\catcode `\#12\catcode `\^12\catcode `\_12\catcode `\%12\relax}%
\providecommand \@@startlink[1]{}%
\providecommand \@@endlink[0]{}%
\providecommand \url  [0]{\begingroup\@sanitize@url \@url }%
\providecommand \@url [1]{\endgroup\@href {#1}{\urlprefix }}%
\providecommand \urlprefix  [0]{URL }%
\providecommand \Eprint [0]{\href }%
\providecommand \doibase [0]{http://dx.doi.org/}%
\providecommand \selectlanguage [0]{\@gobble}%
\providecommand \bibinfo  [0]{\@secondoftwo}%
\providecommand \bibfield  [0]{\@secondoftwo}%
\providecommand \translation [1]{[#1]}%
\providecommand \BibitemOpen [0]{}%
\providecommand \bibitemStop [0]{}%
\providecommand \bibitemNoStop [0]{.\EOS\space}%
\providecommand \EOS [0]{\spacefactor3000\relax}%
\providecommand \BibitemShut  [1]{\csname bibitem#1\endcsname}%
\let\auto@bib@innerbib\@empty
\bibitem [{\citenamefont {Andreev}\ and\ \citenamefont
  {Bashkin}(1975)}]{AndreevBashkin1975}%
  \BibitemOpen
  \bibfield  {author} {\bibinfo {author} {\bibfnamefont {A.}~\bibnamefont
  {Andreev}}\ and\ \bibinfo {author} {\bibfnamefont {E.}~\bibnamefont
  {Bashkin}},\ }\href {http://jetp.ac.ru/cgi-bin/e/index/e/42/1/p164?a=list}
  {\bibfield  {journal} {\bibinfo  {journal} {Zh. Eksp. Teor. Fiz.}\ }\textbf
  {\bibinfo {volume} {69}},\ \bibinfo {pages} {319} (\bibinfo {year}
  {1975})}\BibitemShut {NoStop}%
\bibitem [{\citenamefont {Leggett}(1975)}]{Leggett1975}%
  \BibitemOpen
  \bibfield  {author} {\bibinfo {author} {\bibfnamefont {A.~J.}\ \bibnamefont
  {Leggett}},\ }\href {\doibase 10.1103/RevModPhys.47.331} {\bibfield
  {journal} {\bibinfo  {journal} {Rev. Mod. Phys.}\ }\textbf {\bibinfo {volume}
  {47}},\ \bibinfo {pages} {331} (\bibinfo {year} {1975})}\BibitemShut
  {NoStop}%
\bibitem [{\citenamefont {Sjöberg}(1976)}]{SJOBERG1976}%
  \BibitemOpen
  \bibfield  {author} {\bibinfo {author} {\bibfnamefont {O.}~\bibnamefont
  {Sjöberg}},\ }\href {\doibase https://doi.org/10.1016/0375-9474(76)90558-3}
  {\bibfield  {journal} {\bibinfo  {journal} {Nuclear Physics A}\ }\textbf
  {\bibinfo {volume} {265}},\ \bibinfo {pages} {511 } (\bibinfo {year}
  {1976})}\BibitemShut {NoStop}%
\bibitem [{\citenamefont {Chamel}(2008)}]{Chamel2008}%
  \BibitemOpen
  \bibfield  {author} {\bibinfo {author} {\bibfnamefont {N.}~\bibnamefont
  {Chamel}},\ }\href {\doibase 10.1111/j.1365-2966.2008.13426.x} {\bibfield
  {journal} {\bibinfo  {journal} {Monthly Notices of the Royal Astronomical
  Society}\ }\textbf {\bibinfo {volume} {388}},\ \bibinfo {pages} {737}
  (\bibinfo {year} {2008})}\BibitemShut {NoStop}%
\bibitem [{\citenamefont {{Alpar}}\ \emph {et~al.}(1984)\citenamefont
  {{Alpar}}, \citenamefont {{Langer}},\ and\ \citenamefont
  {{Sauls}}}]{Alpar1984}%
  \BibitemOpen
  \bibfield  {author} {\bibinfo {author} {\bibfnamefont {M.~A.}\ \bibnamefont
  {{Alpar}}}, \bibinfo {author} {\bibfnamefont {S.~A.}\ \bibnamefont
  {{Langer}}}, \ and\ \bibinfo {author} {\bibfnamefont {J.~A.}\ \bibnamefont
  {{Sauls}}},\ }\href {\doibase 10.1086/162232} {\bibfield  {journal} {\bibinfo
   {journal} {\apj}\ }\textbf {\bibinfo {volume} {282}},\ \bibinfo {pages}
  {533} (\bibinfo {year} {1984})}\BibitemShut {NoStop}%
\bibitem [{\citenamefont {Alford}\ and\ \citenamefont
  {Good}(2008)}]{Alford2008}%
  \BibitemOpen
  \bibfield  {author} {\bibinfo {author} {\bibfnamefont {M.~G.}\ \bibnamefont
  {Alford}}\ and\ \bibinfo {author} {\bibfnamefont {G.}~\bibnamefont {Good}},\
  }\href {\doibase 10.1103/PhysRevB.78.024510} {\bibfield  {journal} {\bibinfo
  {journal} {Phys. Rev. B}\ }\textbf {\bibinfo {volume} {78}},\ \bibinfo
  {pages} {024510} (\bibinfo {year} {2008})}\BibitemShut {NoStop}%
\bibitem [{\citenamefont {Babaev}(2009)}]{Babaev2009neutron}%
  \BibitemOpen
  \bibfield  {author} {\bibinfo {author} {\bibfnamefont {E.}~\bibnamefont
  {Babaev}},\ }\href {\doibase 10.1103/PhysRevLett.103.231101} {\bibfield
  {journal} {\bibinfo  {journal} {Phys. Rev. Lett.}\ }\textbf {\bibinfo
  {volume} {103}},\ \bibinfo {pages} {231101} (\bibinfo {year}
  {2009})}\BibitemShut {NoStop}%
\bibitem [{\citenamefont {Kuklov}\ and\ \citenamefont
  {Svistunov}(2003)}]{Kuklov2003}%
  \BibitemOpen
  \bibfield  {author} {\bibinfo {author} {\bibfnamefont {A.~B.}\ \bibnamefont
  {Kuklov}}\ and\ \bibinfo {author} {\bibfnamefont {B.~V.}\ \bibnamefont
  {Svistunov}},\ }\href {\doibase 10.1103/PhysRevLett.90.100401} {\bibfield
  {journal} {\bibinfo  {journal} {Phys. Rev. Lett.}\ }\textbf {\bibinfo
  {volume} {90}},\ \bibinfo {pages} {100401} (\bibinfo {year}
  {2003})}\BibitemShut {NoStop}%
\bibitem [{\citenamefont {Kuklov}\ \emph
  {et~al.}(2004{\natexlab{a}})\citenamefont {Kuklov}, \citenamefont
  {Prokof'ev},\ and\ \citenamefont {Svistunov}}]{Kuklov20042}%
  \BibitemOpen
  \bibfield  {author} {\bibinfo {author} {\bibfnamefont {A.}~\bibnamefont
  {Kuklov}}, \bibinfo {author} {\bibfnamefont {N.}~\bibnamefont {Prokof'ev}}, \
  and\ \bibinfo {author} {\bibfnamefont {B.}~\bibnamefont {Svistunov}},\ }\href
  {\doibase 10.1103/PhysRevLett.92.030403} {\bibfield  {journal} {\bibinfo
  {journal} {Phys. Rev. Lett.}\ }\textbf {\bibinfo {volume} {92}},\ \bibinfo
  {pages} {030403} (\bibinfo {year} {2004}{\natexlab{a}})}\BibitemShut
  {NoStop}%
\bibitem [{\citenamefont {Kuklov}\ \emph
  {et~al.}(2004{\natexlab{b}})\citenamefont {Kuklov}, \citenamefont
  {Prokof'ev},\ and\ \citenamefont {Svistunov}}]{Kuklov2004}%
  \BibitemOpen
  \bibfield  {author} {\bibinfo {author} {\bibfnamefont {A.}~\bibnamefont
  {Kuklov}}, \bibinfo {author} {\bibfnamefont {N.}~\bibnamefont {Prokof'ev}}, \
  and\ \bibinfo {author} {\bibfnamefont {B.}~\bibnamefont {Svistunov}},\ }\href
  {\doibase 10.1103/PhysRevLett.92.050402} {\bibfield  {journal} {\bibinfo
  {journal} {Phys. Rev. Lett.}\ }\textbf {\bibinfo {volume} {92}},\ \bibinfo
  {pages} {050402} (\bibinfo {year} {2004}{\natexlab{b}})}\BibitemShut
  {NoStop}%
\bibitem [{\citenamefont {Kuklov}\ \emph {et~al.}(2006)\citenamefont {Kuklov},
  \citenamefont {Prokof’ev}, \citenamefont {Svistunov},\ and\ \citenamefont
  {Troyer}}]{Kuklov2006}%
  \BibitemOpen
  \bibfield  {author} {\bibinfo {author} {\bibfnamefont {A.}~\bibnamefont
  {Kuklov}}, \bibinfo {author} {\bibfnamefont {N.}~\bibnamefont {Prokof’ev}},
  \bibinfo {author} {\bibfnamefont {B.}~\bibnamefont {Svistunov}}, \ and\
  \bibinfo {author} {\bibfnamefont {M.}~\bibnamefont {Troyer}},\ }\href
  {\doibase https://doi.org/10.1016/j.aop.2006.04.007} {\bibfield  {journal}
  {\bibinfo  {journal} {Annals of Physics}\ }\textbf {\bibinfo {volume}
  {321}},\ \bibinfo {pages} {1602 } (\bibinfo {year} {2006})},\ \bibinfo {note}
  {july 2006 Special Issue}\BibitemShut {NoStop}%
\bibitem [{\citenamefont {Herland}\ \emph {et~al.}(2010)\citenamefont
  {Herland}, \citenamefont {Babaev},\ and\ \citenamefont
  {Sudb\o{}}}]{Herland2010}%
  \BibitemOpen
  \bibfield  {author} {\bibinfo {author} {\bibfnamefont {E.~V.}\ \bibnamefont
  {Herland}}, \bibinfo {author} {\bibfnamefont {E.}~\bibnamefont {Babaev}}, \
  and\ \bibinfo {author} {\bibfnamefont {A.}~\bibnamefont {Sudb\o{}}},\ }\href
  {\doibase 10.1103/PhysRevB.82.134511} {\bibfield  {journal} {\bibinfo
  {journal} {Phys. Rev. B}\ }\textbf {\bibinfo {volume} {82}},\ \bibinfo
  {pages} {134511} (\bibinfo {year} {2010})}\BibitemShut {NoStop}%
\bibitem [{\citenamefont {Dahl}\ \emph
  {et~al.}(2008{\natexlab{a}})\citenamefont {Dahl}, \citenamefont {Babaev},
  \citenamefont {Kragset},\ and\ \citenamefont {Sudb\o{}}}]{Dahl2008}%
  \BibitemOpen
  \bibfield  {author} {\bibinfo {author} {\bibfnamefont {E.~K.}\ \bibnamefont
  {Dahl}}, \bibinfo {author} {\bibfnamefont {E.}~\bibnamefont {Babaev}},
  \bibinfo {author} {\bibfnamefont {S.}~\bibnamefont {Kragset}}, \ and\
  \bibinfo {author} {\bibfnamefont {A.}~\bibnamefont {Sudb\o{}}},\ }\href
  {\doibase 10.1103/PhysRevB.77.144519} {\bibfield  {journal} {\bibinfo
  {journal} {Phys. Rev. B}\ }\textbf {\bibinfo {volume} {77}},\ \bibinfo
  {pages} {144519} (\bibinfo {year} {2008}{\natexlab{a}})}\BibitemShut
  {NoStop}%
\bibitem [{\citenamefont {Greiner}\ \emph {et~al.}(2002)\citenamefont
  {Greiner}, \citenamefont {Mandel}, \citenamefont {Esslinger}, \citenamefont
  {H{\"a}nsch},\ and\ \citenamefont {Bloch}}]{greiner2002quantum}%
  \BibitemOpen
  \bibfield  {author} {\bibinfo {author} {\bibfnamefont {M.}~\bibnamefont
  {Greiner}}, \bibinfo {author} {\bibfnamefont {O.}~\bibnamefont {Mandel}},
  \bibinfo {author} {\bibfnamefont {T.}~\bibnamefont {Esslinger}}, \bibinfo
  {author} {\bibfnamefont {T.~W.}\ \bibnamefont {H{\"a}nsch}}, \ and\ \bibinfo
  {author} {\bibfnamefont {I.}~\bibnamefont {Bloch}},\ }\href@noop {}
  {\bibfield  {journal} {\bibinfo  {journal} {nature}\ }\textbf {\bibinfo
  {volume} {415}},\ \bibinfo {pages} {39} (\bibinfo {year} {2002})}\BibitemShut
  {NoStop}%
\bibitem [{\citenamefont {Bloch}(2005)}]{bloch2005ultracold}%
  \BibitemOpen
  \bibfield  {author} {\bibinfo {author} {\bibfnamefont {I.}~\bibnamefont
  {Bloch}},\ }\href@noop {} {\bibfield  {journal} {\bibinfo  {journal} {Nature
  physics}\ }\textbf {\bibinfo {volume} {1}},\ \bibinfo {pages} {23} (\bibinfo
  {year} {2005})}\BibitemShut {NoStop}%
\bibitem [{\citenamefont {Sellin}\ and\ \citenamefont
  {Babaev}(2018)}]{Sellin2018}%
  \BibitemOpen
  \bibfield  {author} {\bibinfo {author} {\bibfnamefont {K.}~\bibnamefont
  {Sellin}}\ and\ \bibinfo {author} {\bibfnamefont {E.}~\bibnamefont
  {Babaev}},\ }\href {\doibase 10.1103/PhysRevB.97.094517} {\bibfield
  {journal} {\bibinfo  {journal} {Phys. Rev. B}\ }\textbf {\bibinfo {volume}
  {97}},\ \bibinfo {pages} {094517} (\bibinfo {year} {2018})}\BibitemShut
  {NoStop}%
\bibitem [{\citenamefont {S{\"o}yler}\ \emph {et~al.}(2009)\citenamefont
  {S{\"o}yler}, \citenamefont {Capogrosso-Sansone}, \citenamefont {Prokof'ev},\
  and\ \citenamefont {Svistunov}}]{soyler2009sign}%
  \BibitemOpen
  \bibfield  {author} {\bibinfo {author} {\bibfnamefont {{\c{S}}.~G.}\
  \bibnamefont {S{\"o}yler}}, \bibinfo {author} {\bibfnamefont
  {B.}~\bibnamefont {Capogrosso-Sansone}}, \bibinfo {author} {\bibfnamefont
  {N.}~\bibnamefont {Prokof'ev}}, \ and\ \bibinfo {author} {\bibfnamefont
  {B.}~\bibnamefont {Svistunov}},\ }\href@noop {} {\bibfield  {journal}
  {\bibinfo  {journal} {New Journal of Physics}\ }\textbf {\bibinfo {volume}
  {11}},\ \bibinfo {pages} {073036} (\bibinfo {year} {2009})}\BibitemShut
  {NoStop}%
\bibitem [{\citenamefont {Svistunov}\ \emph {et~al.}(2015)\citenamefont
  {Svistunov}, \citenamefont {Babaev},\ and\ \citenamefont
  {Prokofev}}]{EgorBook}%
  \BibitemOpen
  \bibfield  {author} {\bibinfo {author} {\bibfnamefont {B.}~\bibnamefont
  {Svistunov}}, \bibinfo {author} {\bibfnamefont {E.}~\bibnamefont {Babaev}}, \
  and\ \bibinfo {author} {\bibfnamefont {N.~V.}\ \bibnamefont {Prokofev}},\
  }\href {\doibase 10.1201/b18346} {\emph {\bibinfo {title} {Superfluid states
  of matter}}}\ (\bibinfo  {publisher} {CRC Press, Boca Raton, FL},\ \bibinfo
  {year} {2015})\ pp.\ \bibinfo {pages} {1--546}\BibitemShut {NoStop}%
\bibitem [{\citenamefont {Dahl}\ \emph
  {et~al.}(2008{\natexlab{b}})\citenamefont {Dahl}, \citenamefont {Babaev},\
  and\ \citenamefont {Sudb\o{}}}]{Dahl2008_2}%
  \BibitemOpen
  \bibfield  {author} {\bibinfo {author} {\bibfnamefont {E.~K.}\ \bibnamefont
  {Dahl}}, \bibinfo {author} {\bibfnamefont {E.}~\bibnamefont {Babaev}}, \ and\
  \bibinfo {author} {\bibfnamefont {A.}~\bibnamefont {Sudb\o{}}},\ }\href
  {\doibase 10.1103/PhysRevB.78.144510} {\bibfield  {journal} {\bibinfo
  {journal} {Phys. Rev. B}\ }\textbf {\bibinfo {volume} {78}},\ \bibinfo
  {pages} {144510} (\bibinfo {year} {2008}{\natexlab{b}})}\BibitemShut
  {NoStop}%
\bibitem [{\citenamefont {Dahl}\ \emph
  {et~al.}(2008{\natexlab{c}})\citenamefont {Dahl}, \citenamefont {Babaev},\
  and\ \citenamefont {Sudb\o{}}}]{Dahlprl2018}%
  \BibitemOpen
  \bibfield  {author} {\bibinfo {author} {\bibfnamefont {E.~K.}\ \bibnamefont
  {Dahl}}, \bibinfo {author} {\bibfnamefont {E.}~\bibnamefont {Babaev}}, \ and\
  \bibinfo {author} {\bibfnamefont {A.}~\bibnamefont {Sudb\o{}}},\ }\href
  {\doibase 10.1103/PhysRevLett.101.255301} {\bibfield  {journal} {\bibinfo
  {journal} {Phys. Rev. Lett.}\ }\textbf {\bibinfo {volume} {101}},\ \bibinfo
  {pages} {255301} (\bibinfo {year} {2008}{\natexlab{c}})}\BibitemShut
  {NoStop}%
\bibitem [{SI()}]{SI}%
  \BibitemOpen
  \href@noop {} {\enquote {\bibinfo {title} {See supplemental material for
  details on the hamiltonian diagonalization, free energy expansion,
  generalization of the pollock-ceperley formula, worm-algorithm monte carlo
  method, symmetry analysis, and finite size scaling of the effect in strongly
  interacting regime},}\ }\BibitemShut {NoStop}%
\bibitem [{\citenamefont {D'Yakonov}\ and\ \citenamefont
  {Perel}(1971)}]{d1971possibility}%
  \BibitemOpen
  \bibfield  {author} {\bibinfo {author} {\bibfnamefont {M.~I.}\ \bibnamefont
  {D'Yakonov}}\ and\ \bibinfo {author} {\bibfnamefont {V.}~\bibnamefont
  {Perel}},\ }\href@noop {} {\bibfield  {journal} {\bibinfo  {journal} {Soviet
  Journal of Experimental and Theoretical Physics Letters}\ }\textbf {\bibinfo
  {volume} {13}},\ \bibinfo {pages} {467} (\bibinfo {year} {1971})}\BibitemShut
  {NoStop}%
\bibitem [{\citenamefont {Bergeret}\ and\ \citenamefont
  {Tokatly}(2016)}]{Bergeret2016}%
  \BibitemOpen
  \bibfield  {author} {\bibinfo {author} {\bibfnamefont {F.~S.}\ \bibnamefont
  {Bergeret}}\ and\ \bibinfo {author} {\bibfnamefont {I.~V.}\ \bibnamefont
  {Tokatly}},\ }\href {\doibase 10.1103/PhysRevB.94.180502} {\bibfield
  {journal} {\bibinfo  {journal} {Phys. Rev. B}\ }\textbf {\bibinfo {volume}
  {94}},\ \bibinfo {pages} {180502} (\bibinfo {year} {2016})}\BibitemShut
  {NoStop}%
\bibitem [{\citenamefont {Linder}\ \emph {et~al.}(2017)\citenamefont {Linder},
  \citenamefont {Amundsen},\ and\ \citenamefont {Risingg\aa{}rd}}]{Linder2017}%
  \BibitemOpen
  \bibfield  {author} {\bibinfo {author} {\bibfnamefont {J.}~\bibnamefont
  {Linder}}, \bibinfo {author} {\bibfnamefont {M.}~\bibnamefont {Amundsen}}, \
  and\ \bibinfo {author} {\bibfnamefont {V.}~\bibnamefont {Risingg\aa{}rd}},\
  }\href {\doibase 10.1103/PhysRevB.96.094512} {\bibfield  {journal} {\bibinfo
  {journal} {Phys. Rev. B}\ }\textbf {\bibinfo {volume} {96}},\ \bibinfo
  {pages} {094512} (\bibinfo {year} {2017})}\BibitemShut {NoStop}%
\bibitem [{\citenamefont {Fil}\ and\ \citenamefont
  {Shevchenko}(2005)}]{Fil&Shevchenko2005}%
  \BibitemOpen
  \bibfield  {author} {\bibinfo {author} {\bibfnamefont {D.~V.}\ \bibnamefont
  {Fil}}\ and\ \bibinfo {author} {\bibfnamefont {S.~I.}\ \bibnamefont
  {Shevchenko}},\ }\href {\doibase 10.1103/PhysRevA.72.013616} {\bibfield
  {journal} {\bibinfo  {journal} {Phys. Rev. A}\ }\textbf {\bibinfo {volume}
  {72}},\ \bibinfo {pages} {013616} (\bibinfo {year} {2005})}\BibitemShut
  {NoStop}%
\bibitem [{\citenamefont {Linder}\ and\ \citenamefont
  {Sudb\o{}}(2009)}]{Linder&Sudbo2009}%
  \BibitemOpen
  \bibfield  {author} {\bibinfo {author} {\bibfnamefont {J.}~\bibnamefont
  {Linder}}\ and\ \bibinfo {author} {\bibfnamefont {A.}~\bibnamefont
  {Sudb\o{}}},\ }\href {\doibase 10.1103/PhysRevA.79.063610} {\bibfield
  {journal} {\bibinfo  {journal} {Phys. Rev. A}\ }\textbf {\bibinfo {volume}
  {79}},\ \bibinfo {pages} {063610} (\bibinfo {year} {2009})}\BibitemShut
  {NoStop}%
\bibitem [{\citenamefont {Hofer}\ \emph {et~al.}(2012)\citenamefont {Hofer},
  \citenamefont {Bruder},\ and\ \citenamefont {Stojanovi\ifmmode~\acute{c}\else
  \'{c}\fi{}}}]{Hofer1012}%
  \BibitemOpen
  \bibfield  {author} {\bibinfo {author} {\bibfnamefont {P.~P.}\ \bibnamefont
  {Hofer}}, \bibinfo {author} {\bibfnamefont {C.}~\bibnamefont {Bruder}}, \
  and\ \bibinfo {author} {\bibfnamefont {V.~M.}\ \bibnamefont
  {Stojanovi\ifmmode~\acute{c}\else \'{c}\fi{}}},\ }\href {\doibase
  10.1103/PhysRevA.86.033627} {\bibfield  {journal} {\bibinfo  {journal} {Phys.
  Rev. A}\ }\textbf {\bibinfo {volume} {86}},\ \bibinfo {pages} {033627}
  (\bibinfo {year} {2012})}\BibitemShut {NoStop}%
\bibitem [{\citenamefont {Hartman}\ \emph {et~al.}(2018)\citenamefont
  {Hartman}, \citenamefont {Erlandsen},\ and\ \citenamefont
  {Sudb\o{}}}]{Hartman2018}%
  \BibitemOpen
  \bibfield  {author} {\bibinfo {author} {\bibfnamefont {S.}~\bibnamefont
  {Hartman}}, \bibinfo {author} {\bibfnamefont {E.}~\bibnamefont {Erlandsen}},
  \ and\ \bibinfo {author} {\bibfnamefont {A.}~\bibnamefont {Sudb\o{}}},\
  }\href {\doibase 10.1103/PhysRevB.98.024512} {\bibfield  {journal} {\bibinfo
  {journal} {Phys. Rev. B}\ }\textbf {\bibinfo {volume} {98}},\ \bibinfo
  {pages} {024512} (\bibinfo {year} {2018})}\BibitemShut {NoStop}%
\bibitem [{\citenamefont {Nespolo}\ \emph {et~al.}(2017)\citenamefont
  {Nespolo}, \citenamefont {Astrakharchik},\ and\ \citenamefont
  {Recati}}]{nespolo2017andreev}%
  \BibitemOpen
  \bibfield  {author} {\bibinfo {author} {\bibfnamefont {J.}~\bibnamefont
  {Nespolo}}, \bibinfo {author} {\bibfnamefont {G.~E.}\ \bibnamefont
  {Astrakharchik}}, \ and\ \bibinfo {author} {\bibfnamefont {A.}~\bibnamefont
  {Recati}},\ }\href@noop {} {\bibfield  {journal} {\bibinfo  {journal} {New
  Journal of Physics}\ }\textbf {\bibinfo {volume} {19}},\ \bibinfo {pages}
  {125005} (\bibinfo {year} {2017})}\BibitemShut {NoStop}%
\bibitem [{\citenamefont {Prokof'ev}\ \emph {et~al.}(1998)\citenamefont
  {Prokof'ev}, \citenamefont {Svistunov},\ and\ \citenamefont
  {Tupitsyn}}]{Prokofev1998}%
  \BibitemOpen
  \bibfield  {author} {\bibinfo {author} {\bibfnamefont {N.~V.}\ \bibnamefont
  {Prokof'ev}}, \bibinfo {author} {\bibfnamefont {B.~V.}\ \bibnamefont
  {Svistunov}}, \ and\ \bibinfo {author} {\bibfnamefont {I.~S.}\ \bibnamefont
  {Tupitsyn}},\ }\href {\doibase 10.1134/1.558661} {\bibfield  {journal}
  {\bibinfo  {journal} {Journal of Experimental and Theoretical Physics}\
  }\textbf {\bibinfo {volume} {87}},\ \bibinfo {pages} {310} (\bibinfo {year}
  {1998})}\BibitemShut {NoStop}%
\bibitem [{\citenamefont {Capogrosso-Sansone}\ \emph
  {et~al.}(2007)\citenamefont {Capogrosso-Sansone}, \citenamefont {Prokof'ev},\
  and\ \citenamefont {Svistunov}}]{PhysRevB.75.134302}%
  \BibitemOpen
  \bibfield  {author} {\bibinfo {author} {\bibfnamefont {B.}~\bibnamefont
  {Capogrosso-Sansone}}, \bibinfo {author} {\bibfnamefont {N.~V.}\ \bibnamefont
  {Prokof'ev}}, \ and\ \bibinfo {author} {\bibfnamefont {B.~V.}\ \bibnamefont
  {Svistunov}},\ }\href {\doibase 10.1103/PhysRevB.75.134302} {\bibfield
  {journal} {\bibinfo  {journal} {Phys. Rev. B}\ }\textbf {\bibinfo {volume}
  {75}},\ \bibinfo {pages} {134302} (\bibinfo {year} {2007})}\BibitemShut
  {NoStop}%
\bibitem [{\citenamefont {Pollock}\ and\ \citenamefont
  {Ceperley}(1987)}]{PhysRevB.36.8343}%
  \BibitemOpen
  \bibfield  {author} {\bibinfo {author} {\bibfnamefont {E.~L.}\ \bibnamefont
  {Pollock}}\ and\ \bibinfo {author} {\bibfnamefont {D.~M.}\ \bibnamefont
  {Ceperley}},\ }\href {\doibase 10.1103/PhysRevB.36.8343} {\bibfield
  {journal} {\bibinfo  {journal} {Phys. Rev. B}\ }\textbf {\bibinfo {volume}
  {36}},\ \bibinfo {pages} {8343} (\bibinfo {year} {1987})}\BibitemShut
  {NoStop}%
\bibitem [{\citenamefont {Macia}\ \emph {et~al.}(2014)\citenamefont {Macia},
  \citenamefont {Astrakharchik}, \citenamefont {Mazzanti}, \citenamefont
  {Giorgini},\ and\ \citenamefont {Boronat}}]{PhysRevA.90.043623}%
  \BibitemOpen
  \bibfield  {author} {\bibinfo {author} {\bibfnamefont {A.}~\bibnamefont
  {Macia}}, \bibinfo {author} {\bibfnamefont {G.~E.}\ \bibnamefont
  {Astrakharchik}}, \bibinfo {author} {\bibfnamefont {F.}~\bibnamefont
  {Mazzanti}}, \bibinfo {author} {\bibfnamefont {S.}~\bibnamefont {Giorgini}},
  \ and\ \bibinfo {author} {\bibfnamefont {J.}~\bibnamefont {Boronat}},\ }\href
  {\doibase 10.1103/PhysRevA.90.043623} {\bibfield  {journal} {\bibinfo
  {journal} {Phys. Rev. A}\ }\textbf {\bibinfo {volume} {90}},\ \bibinfo
  {pages} {043623} (\bibinfo {year} {2014})}\BibitemShut {NoStop}%
\bibitem [{\citenamefont {Safavi-Naini}\ \emph {et~al.}(2013)\citenamefont
  {Safavi-Naini}, \citenamefont {Söyler}, \citenamefont {Pupillo},
  \citenamefont {Sadeghpour},\ and\ \citenamefont
  {Capogrosso-Sansone}}]{Safavi_Naini_2013}%
  \BibitemOpen
  \bibfield  {author} {\bibinfo {author} {\bibfnamefont {A.}~\bibnamefont
  {Safavi-Naini}}, \bibinfo {author} {\bibfnamefont {{\c{S}}.~G.}\ \bibnamefont
  {Söyler}}, \bibinfo {author} {\bibfnamefont {G.}~\bibnamefont {Pupillo}},
  \bibinfo {author} {\bibfnamefont {H.~R.}\ \bibnamefont {Sadeghpour}}, \ and\
  \bibinfo {author} {\bibfnamefont {B.}~\bibnamefont {Capogrosso-Sansone}},\
  }\href {\doibase 10.1088/1367-2630/15/1/013036} {\bibfield  {journal}
  {\bibinfo  {journal} {New Journal of Physics}\ }\textbf {\bibinfo {volume}
  {15}},\ \bibinfo {pages} {013036} (\bibinfo {year} {2013})}\BibitemShut
  {NoStop}%
\end{thebibliography}%




\setcounter{table}{0}
\setcounter{figure}{0}
\setcounter{equation}{0}
\renewcommand{\thetable}{S\arabic{table}}
\renewcommand\thefigure{S\arabic{figure}}
\renewcommand\theequation{S\arabic{equation}}
\renewcommand{\theHtable}{Supplement.\thetable}
\renewcommand{\theHfigure}{Supplement.\thefigure}
\renewcommand{\theHequation}{Supplement.\theequation}

\begin{titlepage}
\begin{center}
    \Large {\bf Supplemental Material}

    \vspace{1cm}
\end{center}
\end{titlepage}

\section{Diagonalization of the Hamiltonian}
We consider a two-component Bose-Hubbard-type model on a two-dimensional rectangular lattice consisting of $N\times N$ sites with periodic boundary conditions. The Hamiltonian reads
\begin{equation}
    \label{Meth_Hamiltonian}
    \hat{H}
    =
    -
    \sum_{\alpha}
    \sum_{ij}
    t_{ij\alpha} \, \hat b_{i\alpha}^\dagger \hat b_{j\alpha}
    +
    \frac{1}{2}
    \sum_{\alpha\beta}
    \sum_{ij}
    U_{ij\alpha\beta} \, \hat n_{i\alpha} \hat n_{j\beta} \,,
\end{equation}
where $ \alpha, \beta \in \{ a, b \} $ whilst $ i, j $ run over all lattice sites. Here $ \hat b_{i\alpha} $ ($ \hat b^\dagger_{i\alpha} $) is the bosonic annihilation (creation) operator of component $ \alpha $ at site $ i $, and $ \hat n_{i\alpha} = \hat b^\dagger_{i\alpha} \hat b_{i\alpha} $ is the corresponding particle number operator.
We have chosen to restrict the model parameters to
\begin{align}
    t_{ij\alpha}
    &=
    \left\{
    \begin{array}{llll}
        t_{\alpha}^{s}      && \,\,\,\, \text{for} \quad {\bf r}_j - {\bf r}_i = \pm l_s \vec e_s \\
        f_{\alpha}^{\sigma} && \,\,\,\, \text{for} \quad {\bf r}_j - {\bf r}_i = \pm l_x \vec e_x \pm \sigma l_y \vec e_y \\
        0                   && \,\,\,\, \text{otherwise}
   \end{array}
   \right. ,
    \label{Meth_hoppings}
    \\
    U_{ij\alpha\beta}
    &=
    \left\{
    \begin{array}{lllllllllll}
        U_{\alpha\beta}     && \text{for} \quad {\bf r}_j - {\bf r}_i = \vec 0 \\
        V_{\alpha\beta}^s   && \text{for} \quad {\bf r}_j - {\bf r}_i = \pm l_s\vec e_s \\
        0                   && \text{otherwise}
    \end{array}
    \right. ,
    \label{Meth_interactions}
\end{align} 
where $\sigma=\pm1$ and $ \vec r_i $ indicates the $i$\textsuperscript{th} lattice site position, while $ l_s $ and ${\bf e}_s$ are the lattice constant and unit vector in the $s\in \{x,y\}$ direction, respectively.
We switch to the momentum representation by Fourier-transforming the operators $\hat b_{j\alpha}=\sum_{{\bf k}} \hat b_{{\bf k}\alpha}\mathrm{e}^{-\mathrm{i} {\bf k}\cdot {\bf r}_j}/N$
with ${\bf k}=k_x \vec e_x + k_y \vec e_y$.
In the macroscopic limit at a sufficiently low temperature, when each component macroscopically occupy the ${\bf k}= {\bf 0}$ mode, one can replace $ \hat b_{0\alpha}, \hat b_{0\alpha}^\dagger \rightarrow  \big<\hat{b}_{0\alpha}^\dagger\hat{b}_{0\alpha}\big>^{1/2} = N \sqrt{n_{0\alpha}} $ and express the  particle number density as $  n_{\alpha} = n_{0\alpha} + \sum_{{\bf k}\neq 0} \hat b_{{\bf k}\alpha}^\dagger \hat b_{{\bf k}\alpha}/N^2 $.
By omitting terms containing more than two of the remaining $\hat{b}_{\vec k \neq 0 \alpha}^\dagger$ and $\hat{b}_{\vec k \neq 0 \alpha}$, we finally obtain
\begin{equation}
\label{Ham2ndOrderToDiag}
\begin{split}
    \!\!\! \mathcal{\hat{H}}
    &=
    \mathcal{H}_0 
    + \sum_{\alpha}\sum_{{\bf k}\neq0} \varepsilon_{{\bf k}\alpha}\hat{b}_{{\bf k}\alpha}^\dagger \hat{b}_{{\bf k}\alpha}
    \\
    &+
    \sum_{\alpha}\sum_{{\bf k}\neq0} 
    \frac{\mathcal{U}_{{\bf k}\alpha\alpha}}{2}
    \Big(\hat{b}_{{\bf k}\alpha}^\dagger \hat{b}_{-{\bf k}\alpha}^\dagger +\hat{b}_{-{\bf k}\alpha} \hat{b}_{{\bf k}\alpha}+2\hat{b}_{{\bf k}\alpha}^\dagger \hat{b}_{{\bf k}\alpha}\Big)
     \\
    &
    +\sum_{{\bf k}\neq0}
    \mathcal{U}_{{\bf k}ab}
    \Big( \hat{b}_{{\bf k}a}^\dagger \hat{b}_{{\bf k}b}+ \hat{b}_{{\bf k}a}^\dagger \hat{b}_{-{\bf k}b}^\dagger +\text{h.c.} \Big) \,.
\end{split}
\end{equation}
Here $\mathcal{H}_0$ is a constant and introducing $\xi_s=k_sl_s$, $ \widetilde{U}_{\alpha\beta} = U_{\alpha\beta}\sqrt{n_{\alpha}n_{\beta}} $, and $ \widetilde{V}_{\alpha\beta}^{s} = V_{\alpha\beta}^{s}\sqrt{n_{\alpha}n_{\beta}} $ we defined
\begin{align}
    \label{Ukab_pap}
    &\!
    \mathcal{U}_{{\bf k}\alpha\beta}
    = \widetilde{U}_{\alpha\beta}
    +  2 \big[ \widetilde{V}_{\alpha\beta}^x\cos(\xi_x)
    +  \widetilde{V}_{\alpha\beta}^y\cos(\xi_y) \big],
    \\
    \label{Meth_eps}
    &\!\!\!\!
    \varepsilon_{{\bf k}\alpha}
     \!=\!
     \sum_{s}2t_{\alpha}^s[1\!-\!\cos(\xi_s)]\!
     +\!
     \sum_{\sigma}2f_{\alpha}^\sigma  [1\!-\!\cos(\xi_x\!+\!\sigma\xi_y)].
\end{align}
Since a superfluid flow is related to a phase gradient along the lattice, we may incroporate the corresponding superfluid velocity $ \vec v_\alpha = v_\alpha^x \vec e_x + v_\alpha^y \vec e_y$ in Peierls-like factors, i.e., $t_{ij\alpha} \rightarrow t_{ij\alpha} \, \mathrm{e}^{-\mathrm{i}\Delta\phi^{ij}_\alpha}$. Here
\begin{equation}
    \label{eq:Peierls-factors}
    \Delta \phi_\alpha^{ij}
    =
    m_\alpha \int_{\vec r_j}^{\vec r_i} \vec v_\alpha (\vec x) \cdot \diff \vec x
    \,,
\end{equation}
where $m_\alpha$ is the $\alpha$-type particles mass, $ \hbar=1 $, and the integration taken along a straight line between $i$\textsuperscript{th} and $j$\textsuperscript{th} lattice sites. 
Note that since $\Delta\phi_{\alpha}^{ii}=0$, the introduced phase nonuniformity modifies only the kinetic part of the Hamiltonian and can be incorporated by the substitution ${\bf k}\rightarrow {\bf k}-m_\alpha {\bf v}_\alpha$ in Eq.~(\ref{Meth_eps}).
The Hamiltonian (\ref{Ham2ndOrderToDiag}) can be diagonalized 
similarly as in Refs.~[\hyperref[Fil]{1}--\hyperref[Hartman]{4}] 
and cast into the form
\begin{align}
    \hat{\mathcal{H}}=\widetilde{\mathcal{H}}_0
    + 
    \sum_{\vec k \neq \vec 0}
    \sum_{\sigma=\pm}\mathcal{E}_{{\bf k},\sigma}\left(\beta_{{\bf k},\sigma}^\dagger\beta_{{\bf k},\sigma}+\frac{1}{2}\right)
    \,,
    \label{DiagonalizedHam}
\end{align}
where $ \widetilde{\mathcal{H}}_0 $ is a constant
and $\beta_{{\bf k},\pm}$ are bosonic annihilation operators of $\pm$-type quasiparticles. 
To do so, it is convenient to rewrite Eq.~(\ref{Ham2ndOrderToDiag}) into the matrix form 
\begin{align}
\hat{\mathcal{H}}=\mathcal{H}_0+\frac{1}{4}\sum_{\vec k \neq \vec 0}\psi_{\vec k}^\dagger \mathcal{M}_{\vec k} \psi_{\vec k},
\label{Hmatrix}
\end{align}
where the following basis vectors
\begin{align}
\!\!\!
\psi_{\vec k}=[\hat{b}_{\vec k a},
\, \hat{b}_{-\vec k a},
\, \hat{b}_{\vec k b},
\, \hat{b}_{-\vec k b},
\, 
\hat{b}_{\vec k a}^\dagger,
\, \hat{b}_{-\vec k a}^\dagger,
\, \hat{b}_{\vec k b}^\dagger,
\, \hat{b}_{-\vec k b}^\dagger]^T,
\label{vecs}
\end{align}
and $8\times8$ matrix
\be
\mathcal{M}_{\vec{k}}=\left(
\begin{array}{cc}
A_{\vec{k}} & B_{\vec{k}} \\ 
B_{\vec{k}} & A_{\vec{k}}
\end{array}
\right),
\ee
with
\begin{align}
A_{\vec{k}}=\left(
\begin{array}{cccc}
\widetilde{\varepsilon}_{\vec{k}a} & 0 & \mathcal{U}_{\vec{k}ab} & 0 \\
0 & \widetilde{\varepsilon}_{-\vec{k}a} & 0 & \mathcal{U}_{\vec{k}ab} \\
\mathcal{U}_{\vec{k}ab} & 0 & \widetilde{\varepsilon}_{\vec{k}b} & 0 \\
0 & \mathcal{U}_{\vec{k}ab} & 0 & \widetilde{\varepsilon}_{-\vec{k}b}
\end{array}
\right)
,
\\  
B_{\vec{k}}=\left(
\begin{array}{cccc}
0 & \mathcal{U}_{\vec{k}aa} & 0 & \mathcal{U}_{\vec{k}ab}  \\
\mathcal{U}_{\vec{k}aa} & 0 & \mathcal{U}_{\vec{k}ab} & 0  \\
0 & \mathcal{U}_{\vec{k}ab} & 0 & \mathcal{U}_{\vec{k} bb} \\
\mathcal{U}_{\vec{k}ab} & 0 & \mathcal{U}_{\vec{k} bb} & 0
\end{array}
\right),
\label{auxMats}
\end{align}
and $\widetilde{\varepsilon}_{\vec{k}\alpha}=\varepsilon_{\vec{k}\alpha}+\mathcal{U}_{\vec{k}\alpha\alpha}$ are introduced. The Hamiltonian $\hat{\mathcal{H}}$ can be diagonalized via a unitary transformation $\hat{\mathcal{O}}_{\vec k}$ with the new basis vectors $\Psi_{\vec k}=\hat{\mathcal{O}}_{\vec k}^\dagger \psi_{\vec k}$. The requirement that the new operators $\beta_{\vec k \sigma}, \beta_{\vec k \sigma}^\dagger$ stored in $\Psi_{\vec k}$ satisfy the standard bosonic commutation relations leads to $\hat{\mathcal{O}}_{\vec k}^\dagger \check{\sigma}_3 \hat{\mathcal{O}}_{\vec k}=\check{\sigma}_3$, where
\be
\check{\sigma}_3=\left(
\begin{array}{cc}
{\bf 1}_4 & 0 
\\
0 & - {\bf 1}_4
\end{array}
\right),
\ee
with ${\bf 1}_n$ being the $n$-dimensional identity matrix.
In result, $\mathcal{O}_{\vec k}$ diagonalizes $M_{\vec k} = \mathcal{M}_{\vec k}\check{\sigma}_3$ where the corresponding eigenvalues can be determined by solving the characteristic equation $|M_{\vec k}-\mathcal{E}{\bf 1}_8|=0$. Consequently, one finds four eigenenergies $\pm \mathcal{E}_{\vec k, \sigma=\pm}$ where the positive ones describe quasiparticle excitations (see also~[\hyperref[Fil]{1}--\hyperref[Hartman]{4}]).

\section{Free energy expansion}
The free-energy density of the system at temperature $T=0\mathrm{K}$ is given by an expectation value of the Hamiltonian (\ref{DiagonalizedHam}) in a state free of quasiparticle excitations~[\hyperref[Fil]{1},\hyperref[Hofer]{3},\hyperref[Hartman]{4}]. Since the superfluid velocities are assumed to be small, the eigenenergies  $\mathcal{E}_{{\bf k},\sigma}$  can be determined by expansion 
in terms of small ${\bf v}_a$ and ${\bf v}_b$, where in the zeroth order 
$\mathcal{E}_{{\bf k},\pm}^{(0)} = \sqrt{ (Q_{{\bf k}a}+Q_{{\bf k}b}\pm\sqrt{\Gamma_{{\bf k}} }\,)/2 }  $,
with 
$Q_{{\bf k}\alpha}=\varepsilon_{{\bf k}\alpha}(\varepsilon_{{\bf k}\alpha}+2\,\mathcal{U}_{{\bf k}\alpha})$ and 
$\Gamma_{{\bf k}}= (Q_{{\bf k}a}-Q_{{\bf k}b} )^2 +16\, \varepsilon_{{\bf k}a}\varepsilon_{{\bf k}b}\, \mathcal{U}_{{\bf k}ab}^2 $,
see also Refs.~
[\hyperref[Fil]{1}--\hyperref[Hartman]{4}]
The subsequent higher order terms in the expansion can be found by expanding $\varepsilon_{\vec k \alpha}$ with $\vec k \rightarrow \vec k -m_\alpha \vec v _\alpha$ in terms of small $\vec v_\alpha$ and solving the characteristic equation order by order.
Note, that in order to prevent spatial collapse 
or separation between atomic clouds, the interactions have to satisfy the relation $\mathcal{U}_{{\bf k}a}\mathcal{U}_{{\bf k}b} >\mathcal{U}_{{\bf k}ab}^2$.

For small superfluid velocities the free-energy density can---up to a constant---be cast into the following form
\begin{equation}
    \label{eq:free-energy-density}
     f
    =
    \frac{1}{2}
    \sum_{\alpha\beta}
    \sum_{ij}
    \rho_{\alpha\beta}^{ij}
    v_{\alpha}^i v_{\beta}^j
    =
    \frac 1 2
    \sum_{\alpha\beta}
    \vec v_\alpha^T \rho_{\alpha\beta} \vec v_\beta,
\end{equation}
where 
$ \rho_{\alpha\beta} \definition \sum_{ij} \rho_{\alpha\beta}^{ij} \vec e_i \vec e_j^T $, and 
the drag-related densities are found to be given by
\begin{align}
    \rho_{ab}^{ij}
    =
    \frac{m_a m_b }{N^2} \frac{l_i l_j}{ l_x l_y} \sum_{{\bf k}\neq \vec 0} G({\bf k}) (\partial_{\xi_i}\varepsilon_{{\bf k}a}) (\partial_{\xi_j}\varepsilon_{{\bf k}b}),
    \label{tensordragcoeffs1}
\end{align}
with $G({\bf k})=2\,\varepsilon_{{\bf k}a}\varepsilon_{{\bf k}b} \,\mathcal{U}_{{\bf k}ab}^2  
/\mathcal{E}_{{\bf k},+}^{(0)}\mathcal{E}_{{\bf k},-}^{(0)} 
\big(\mathcal{E}_{{\bf k},+}^{(0)}+\mathcal{E}_{{\bf k},-}^{(0)}\big)^{3}
$,
which vanishes when $\mathcal{U}_{{\bf k}ab}\rightarrow 0$. 
Note that both $G(\vec k)$ and $\varepsilon_{\vec k \alpha}$ can be viewed as functions of $\xi_s$ which is $l_s$ independent. Therefore, the values of Andreev-Bashkin terms $\rho_{ab}^{ii}$ can be simply modified by adjusting the ratio $l_x/l_y$.    
The drag-related elements (\ref{tensordragcoeffs1}) are calculated in the thermodynamic limit where we transit from summation to integration, $\mathcal{V}^{-1}\sum_{\vec k \neq \vec 0} \mathcal{I}(\vec k)   \rightarrow   (2\pi)^{-2}\int_{\text{1BZ}} \mathcal{I}(\vec k) \mathrm{d}^2 k$, 
noting that the integrand $ \mathcal{I}(\vec k) $ vanishes for ${\bf k}= \vec 0$.

\section{Generalization of Pollock-Ceperley Equation}

We will consider a $ n $-component superfluid inhabiting a $ d $-dimensional lattice with periodic boundary conditions. The set of lattice vectors $\{\vec a_i\}_{i=1,\ldots,d}$ need in general not be orthonormal, i.e., $ \vec a_i \cdot \vec a_j \neq \delta_{ij} $ and $ | \vec a_i | = l_i \neq 1 $, where $ l_i $ are the corresponding lattice constants.
The number of lattice sites along the direction of the $i$\textsuperscript{th} lattice vector is given by $ N_i $, and the corresponding side length is therefore $ L_i = N_i l_i $.
Introducing the matrices
$ \mathcal N \definition \sum_{i} N_i \vec e_i \vec e_i^T $ and $ \mathcal M \definition \sum_{i} \vec a_i \vec e_i^T $, where $\{\vec e _i\}_{i=1,\ldots,d}$ is a set of orthonormal coordinate vectors (here $ \vec e_1 = \vec e_x $, $ \vec e_2 = \vec e_y $, etc.), the lattice volume may be expressed as $ \mathcal V = \text{det} (\mathcal N) |\text{det}(\mathcal M)| $.

Following the derivation in~[\hyperref[Sellin]{5}], we set the superfluid components in motion via a component-dependent Galilean transformation. Compared to a motionless system, this will introduce Peierels phase factors in the hopping amplitudes, i.e., $ t_\alpha^{ij} \rightarrow t_\alpha^{ij} e^{-\iu \Delta \phi_\alpha^{ij}} $, with $ \Delta \phi_\alpha^{ij} $ given by \eqref{eq:Peierls-factors}.
For the purpose of this derivation, it is sufficient to consider a uniform velocity field such that $ \Delta \phi_\alpha^{ij} = m_\alpha \vec v_\alpha \cdot (\vec r_i - \vec r_j) $. Due to the periodic boundary conditions, each term in the partition function gains a factor $ \exp[-\iu m_\alpha \vec v_\alpha \cdot ( \pm N_i \vec a_i)] $ for each $\alpha$-component particle crossing the boundary in the $ \pm \vec{a}_i/|\vec a_i| $ direction.
Introducing the winding number $ w_\alpha^i \in \mathbb Z $, as the flux of type $ \alpha $ particles through the boundary perpendicular to $ \vec a_i $, the net phase factor may be expressed as
$ \mathrm{exp}\big(-\iu \sum_{\alpha} m_\alpha \vec v_\alpha^T \mathcal M \mathcal N \vec w_\alpha \big) \definition \mathrm{exp}[\iu \theta (\vec v_\alpha, \vec w_\alpha)] $
where $ \sum_i w_\alpha^i \vec a_i = \mathcal M \sum_i w_\alpha^i \vec e_i = \mathcal M \vec w_\alpha $. Next, we decompose the partition function $Z(\vec v_\alpha)$ in terms of fixed-winding-number partition functions $ Z_{\{ \vec w_\alpha \}} $, i.e., $Z(\vec v_\alpha)=\sum_{\{\vec w_\alpha\}} Z_{\{ \vec w_\alpha \}} e^{\iu \theta (\vec v_\alpha, \vec w_\alpha)}$.
In this notation, the partition function of the motionless system corresponds to $ Z(\vec v_\alpha = \vec 0) $, and the ratio $ Z(\vec v_\alpha) / Z(\vec 0) $ is identified to be the expectation value of the phase factor, namely $ \langle e^{\iu \theta (\vec v_\alpha, \vec w_\alpha)} \rangle $. For small velocities $ \vec v_\alpha $, this quantity may be approximated as
\begin{equation}
    \!\!\!\!
    \label{eq:phase-factor-expansion-small-v}
    \resizebox{.887\hsize}{!}{$%
        \langle e^{\iu \theta (\vec v_\alpha, \vec w_\alpha)} \rangle 
        \! \simeq \!
        1 \! - \!
        \sum \limits_{\alpha\beta}
        \frac{m_\alpha m_\beta}{2}
        \,
        \vec v_\alpha^T
        \mathcal M \mathcal N
        W_{\alpha\beta}
        \mathcal N^T \mathcal M^T
        \vec v_\beta \,,
    $}
\end{equation}
where the winding-number correlation tensor $ W_{\alpha\beta} \definition \langle \vec w_\alpha \vec w_\beta^T \rangle $ has been introduced. The linear terms $ \langle w_\alpha^i \rangle$ vanish since the microscopic model is assumed to be invariant under a parity inversion.

In addition, the ratio $ Z(\vec v_\alpha) / Z(\vec 0) = e^{- \beta \Delta F} $ is also related to the free-energy difference $ \Delta F = F(\vec v_\alpha) - F(\vec 0) $ of having set the system in motion. Here $ \beta $ is the inverse temperature, and due to the previous assumption of uniform velocity fields, the free-energy difference may be expressed as the free-energy density \eqref{eq:free-energy-density} times the volume, i.e., $ \Delta F = \mathcal V f $. In the limit of small velocities $ \vec v_\alpha $, the free-energy exponential may be accurately approximated by the two first terms of its series expansion,
\begin{equation}
    \label{eq:free-energy-expansion-small-v}
    e^{- \beta \mathcal V f}
    \simeq
    1 - \beta \mathcal V f
    =
    1 - \beta \mathcal V \, \frac 1 2 \sum_{\alpha\beta} \vec v_\alpha^T \rho_{\alpha\beta} \vec v_\beta \,.
\end{equation}
Combining Eq.~\eqref{eq:phase-factor-expansion-small-v} and Eq.~\eqref{eq:free-energy-expansion-small-v}, we obtain
\begin{equation}
    \label{eq:Pollock-Cepereley}
    \rho_{\alpha\beta}
    =
    \frac{m_\alpha m_\beta}{\beta \mathcal V}
    \mathcal M \mathcal N W_{\alpha\beta} \mathcal N^T \mathcal M^T \,,
\end{equation}
which generalizes the relation which was first derived in~[\hyperref[Pollock]{6}].

We find that in the case of a $ N \times N $ rectangular lattice, the drag coefficients are given by
\begin{equation}
    \label{eq:Pollock-Cepereley-rectangular}
    \rho_{ab}^{ij}
    =
    \frac{m_a m_b}{\beta}
    \frac{l_i l_j}{l_x l_y}
    \langle w_a^i w_b^j \rangle \,,
\end{equation}
which is an analog of Eq.~\eqref{tensordragcoeffs1} but here valid also in the strongly interacting regime.

\section{Worm-algorithm Monte Carlo}

In this Letter, we use continuous-time worm-algorithm Monte Carlo~[\hyperref[Prokof]{7}] to extract completely unbiased winding-number correlations for the two-component Bose-Hubbard type model, described in the main text.
Worm-algorithm Monte Carlo efficiently samples diagonal elements of the density matrix at thermal equilibrium as world-line configurations in real space and imaginary time. This is achieved by combining the partition function sector with the Green’s function sector, so that, in order to go from one partition function world-line configuration to another, one needs to pass through the Green’s function sector.

The $ d+1 $ dimensional system containing the world-lines is periodic in the spatial dimensions due to imposed periodic boundary conditions. It is also periodic in imaginary time to reflect the cyclic property of the trace, present in the density matrix being sampled.
We use separate worms for the two components, which are able to change their winding numbers $ w_\alpha^i $ by winding in spatial dimensions.
In contrast, the winding number in imaginary time determines the total particle number, and since we here consider a fixed number of particles, the latter type of winding is forbidden.

\section{Symmetry Considerations}
The novel dissipationless transport phenomenon is especially transparent when the representation of the inter-species interaction of Eq.~\eqref{eq:free-energy-density}
reads
\begin{equation}
    \label{eq:free-energy-denisty-interaction}
    f_{ab} = \rho_\parallel \vec v_a \cdot \vec v_b + \rho_\perp (v_a^x v_b^y - v_a^y v_b^x) \,.
\end{equation}
Here 
$ \rho_\parallel \definition (\rho_{ab}^{xx}+\rho_{ab}^{yy})/2 $ and $ \rho_\perp \definition (\rho_{ab}^{xy}-\rho_{ab}^{yx}) / 2 $ are coordinate system independent drag-coefficients.
To achieve this, we first consider rotation and reflection transformations, which in two dimensions can be represented by the orthogonal matrices
\begin{equation}
    \mathcal R_1
    =
    \begin{pmatrix}
        c_\theta & - s_\theta \\
        s_\theta &   c_\theta
    \end{pmatrix} 
    \quad
    \text{and}
    \quad
    \mathcal R_2
    =
    \begin{pmatrix}
        c_{2\varphi} & s_{2\varphi} \\
        s_{2\varphi} &  -c_{2\varphi}
    \end{pmatrix} 
    ,
\end{equation}
respectively. Here $ c_\phi = \cos \phi $ and $ s_\phi = \sin \phi $, so that $\mathcal R_1$ rotates the system by an angle $\theta$, whilst $\mathcal R_2$ reflects against a line making the angle $\varphi$ with the $x$-axis.
Since matrices transform according to $ M' = \mathcal{R}_k M \mathcal{R}_k^{T} $ ($ k \in \{1,2\} $), drag coefficients of the superfluid stiffness tensor will in the rotated system $ {\bf r}' = \mathcal{R}_{k} {\bf r} $ be given by
\begin{align}
    & \rho_{ab}^{x'x'}+\rho_{ab}^{y'y'}
    = \rho_{ab}^{xx}+\rho_{ab}^{yy}
      \equiv 2\, \rho_\parallel \,,
    \label{sigm0}
    \\
    & \rho_{ab}^{x'y'}-\rho_{ab}^{y'x'}
    = \rho_{ab}^{xy}-\rho_{ab}^{yx}
      \equiv 2\, \rho_\perp \,,
    \label{delt0}
    \\
    & \rho_{ab}^{x'x'}-\rho_{ab}^{y'y'}
    = (\rho_{ab}^{xx}-\rho_{ab}^{yy}) c_{2\theta}
    - (\rho_{ab}^{xy}+\rho_{ab}^{yx}) s_{2\theta} \,,
    \label{minus0}
    \\
    & \rho_{ab}^{x'y'}+\rho_{ab}^{y'x'}
    = (\rho_{ab}^{xy}+\rho_{ab}^{yx}) c_{2\theta}
    + (\rho_{ab}^{xx}-\rho_{ab}^{yy}) s_{2\theta} \,.
    \label{plus0}
\end{align}
From this we note that $ \rho_\parallel $ and $ \rho_\perp $ indeed are invariant under rotations, but more importantly that we need $ \rho_{ab}^{xx} = \rho_{ab}^{yy} $ and $ \rho_{ab}^{xy} = - \rho_{ab}^{yx} $ to consistently cancel other contributions to $ f_{ab} $. Note, that since the model is homogeneous, the superfluid stiffness tensor and its components are not affected by translation.

The explicit dependence of $ \rho_{ab}^{ii} $ on $ l_i $, as revealed by Eq.~\eqref{tensordragcoeffs1} and Eq.~\eqref{eq:Pollock-Cepereley-rectangular}, makes it possible to cancel $ \rho_{ab}^{xx} - \rho_{ab}^{yy} $ by scaling the ratio $ l_x/l_y $ by $ \sqrt{\rho_{ab}^{yy}/\rho_{ab}^{xx}} $, as long as $ \rho_{ab}^{yy} $ and $ \rho_{ab}^{xx} $ are nonzero and of the same sign.
However, to fix $ \rho_{ab}^{xy} + \rho_{ab}^{yx} = 0 $, we will consider a reflection in the $x$- or $y$-axis ($ 2\varphi = \pi n \,, n \in \mathbb Z $), combined with a component exchange operation, i.e., $ a \leftrightarrow b $. The former of the two operations yield
$ \rho_{ab}^{xx} \pm \rho_{ab}^{yy} \rightarrow \rho_{ab}^{xx} \pm \rho_{ab}^{yy} $ and
$ \rho_{ab}^{xy} \pm \rho_{ab}^{yx} \rightarrow -(\rho_{ab}^{xy} \pm \rho_{ab}^{yx}) $,
but in combination with the latter we find
$ \rho_{ab}^{xx} \pm \rho_{ab}^{yy} \rightarrow 
\rho_{ab}^{xx} \pm \rho_{ab}^{yy} $ and
$ \rho_{ab}^{xy} \pm \rho_{ab}^{yx} \rightarrow 
\mp (\rho_{ab}^{xy} \pm \rho_{ab}^{yx}) $. Hence $ \rho_\parallel $, $ \rho_\perp $, and $ \rho_{ab}^{xx} + \rho_{ab}^{yy} $ are invariant with respect to this transformation, whereas the sign of $ \rho_{ab}^{xy} + \rho_{ab}^{yx} $ is negated.
From the plain reflection we further observe that it is crucial that the system breaks reflection symmetry in both the $ x $- and $ y $-direction as the vector-drag coefficient otherwise will vanish.
Thus, by incorporating these symmetries into the model in addition to adjusting the $ l_x / l_y $ ratio, we obtain the desired form of the interacting part of the free-energy density \eqref{eq:free-energy-denisty-interaction}.

\section{Lattice asymmetry}

\begin{figure*}[t!]
    \includegraphics[width=0.93\textwidth]{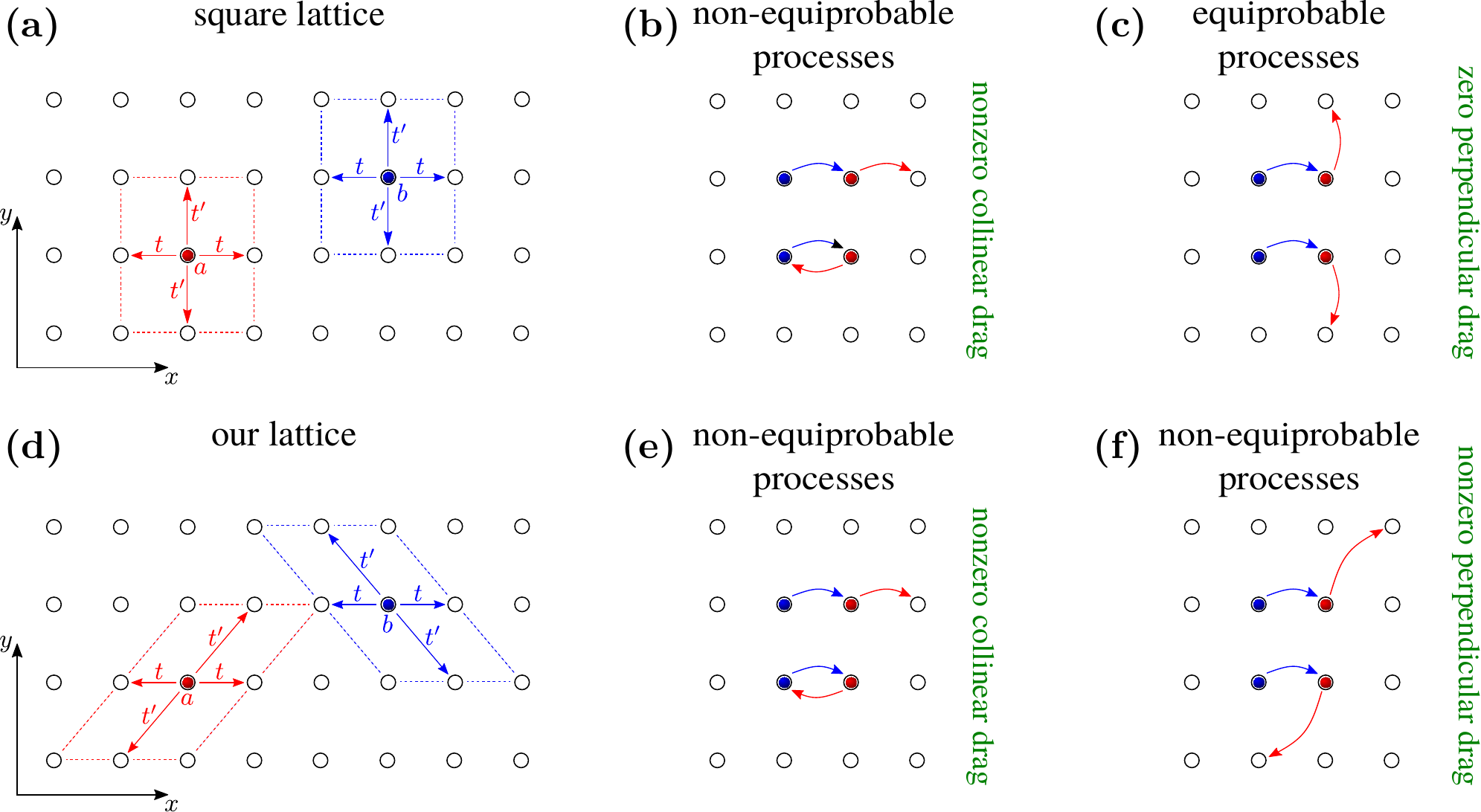}
    \caption{
        A cartoon illustrating the elementary hopping processes 
        of an ordinary square lattice featuring the Andreev-Bashkin effect but zero vector drag {\bf (upper row)},
        and our choice of lattice geometry resulting in the presence of both phenomena {\bf (lower row)}.
    }\vspace{-0.35cm}
    \label{cart_lab}
\end{figure*}

We can make some illustrative remarks here which should not be mistaken for a rigorous description of the phenomena.
The ordinary drag phenomenon arises when
elementary collinear hopping processes on a  lattice,
shown in Fig.~\ref{cart_lab}(b),
are not equally likely.
However, in such a situation the perpendicular component of the drag vanishes on symmetry grounds: elementary perpendicular hopping processes, i.e., the first particle hops to the right ``pushing'' the second particle up or down, are equiprobable, Fig.~\ref{cart_lab}(c).

Unlike the traditional symmetric setting for which the vector drag vanishes, our system is asymmetric 
in respect to interaction-stimulated perpendicular hopping, Fig.~\ref{cart_lab}(d). 
This results in elementary non-collinear hopping processes being non-equiprobable,
Fig.~\ref{cart_lab}(e,f). 
The asymmetry of the elementary processes leaves the imprint
on the superfluid hydrodynamics resulting
in a nonzero vector-drag coefficient $ \rho_\perp $.

\vspace{-0.35cm}

\section{Finite size scaling}
\begin{figure}[h!]
  \centering
  \includegraphics[width=1.\columnwidth]{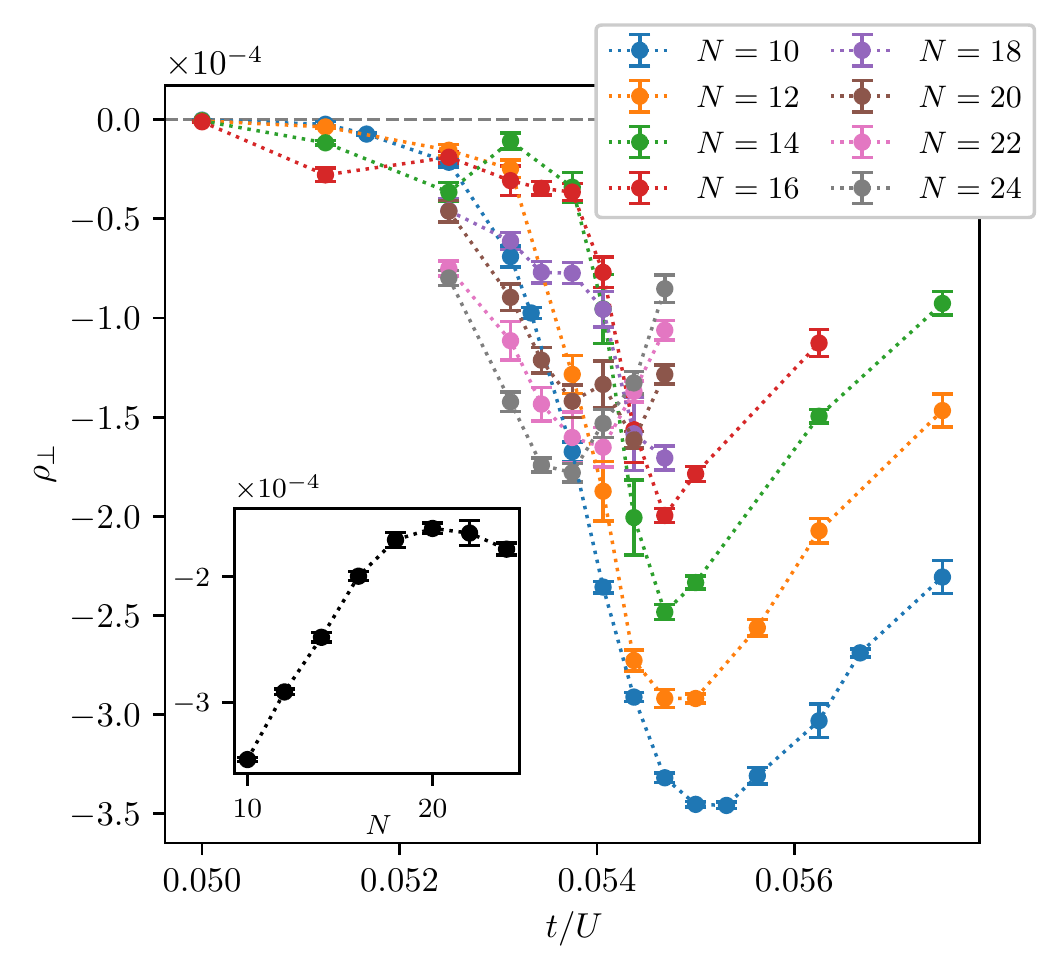}
  \vspace{-0.75cm}
  \caption{
   The vector-drag coefficient $ \rho_\perp $ in the absence of nearest-neighbor interaction, i.e., $ V=V'=0 $, for different system sizes $ N $, as a function of $ t/U $. The inset shows the peak value for each system size, which shifts towards lower $ t/U $ with increased $ N $. Initially the peak values decrease with increased $ N $, but at $ N = 20 $ the trend  changes: the   peak value starts   increasing  with increased $ N $.
    The remaining parameters used are $ t = t' $, $ U'/U = 0.9 $, $ \beta = N/t $ , and $ n_a = n_b = 1/2 $.
  }
  \vspace{-0.5cm}
  \label{fig:rho_perp_vs_N}
\end{figure}

Here we investigate how the vector-drag coefficient $ \rho_\perp $ scale with the system size $ N $, in the absence of nearest-neighbor interactions, i.e., $ V = V'=0 $.
From Fig.~\ref{fig:rho_perp_vs_N} we notice a shift in the peak position towards lower values of $ t/U $ when $ N $ is increased, which initially is accompanied by a decrease in the peak value. However, at $ N = 20 $, the trend changes and the peak value starts growing when further increasing $ N $.
This suggests that the vector drag in the model with exclusively on-site interactions
is not a finite-size effect.

\section{References}



\label{Fil}
\noindent
[1] D. V. Fil and S. I. Shevchenko, 
\href{\doibase 10.1103/PhysRevA.72.013616}{Phys. Rev. A {\bf 72}, 013616 (2005)}
\vspace{0.1cm}

\label{Linder}
\noindent
[2] %
J. Linder and A. Sudb\o{}, \href {\doibase 10.1103/PhysRevA.79.063610}{Phys. Rev. A79, 063610 (2009)}.
\vspace{0.1cm}

\label{Hofer}
\noindent
[3]
P. P. Hofer, C. Bruder,  and V. M. Stojanovi\ifmmode \acute{c}\else \'{c}\fi{}, 
\href{\doibase 10.1103/PhysRevA.86.033627}{Phys. Rev. A {\bf 86}, 033627 (2012)}.
\vspace{0.1cm}

\label{Hartman}
\noindent
[4]  S. Hartman, E. Erlandsen,   and A. Sudb\o{}, Phys. Rev. B {\bf 98}, 024512 (2018).
\href {\doibase 10.1103/PhysRevB.98.024512} {Phys. Rev. B {\bf 98}, 024512 (2018)}
\vspace{0.1cm}

\label{Sellin}
\noindent
[5]  K. Sellin and E. Babaev, \href{\doibase 10.1103/PhysRevB.97.094517}{Phys. Rev. B {\bf 97}, 094517 (2018)}.
\vspace{0.1cm}

\label{Pollock}
\noindent
[6]  E. L. Pollock and D. M. Ceperley, \href{\doibase 10.1103/PhysRevB.36.8343}{Phys. Rev. B {\bf 36}, 8343 (1987)}.
\vspace{0.1cm}

\label{Prokof}
\noindent
[7]  N.  V.  Prokof’ev,  B.  V.  Svistunov,    and  I.  S.  Tupitsyn,  \href{\doibase 10.1134/1.558661}{Journal of Experimental and Theoretical Physics {\bf 87}, 310 (1998)}.


\end{document}